\documentclass[
  pra,
  twocolumn,
  showpacs,
  amsmath,
  amssymb,
  superscriptaddress, 
]{revtex4}
\usepackage[colorlinks]{hyperref}
\usepackage{bm}
\usepackage{graphicx}

\def\be{\begin{equation}}
\def\ee{\end{equation}}
\def\bea{\begin{eqnarray}}
\def\eea{\end{eqnarray}}

\begin{document}

\title{Energy transport in the Anderson insulator}

\author{D.\ B.\ Gutman}
\affiliation{Department of Physics, Bar Ilan University, Ramat Gan 52900,
Israel }

\author{I.V.Protopopov}
\affiliation{
 Institut f\"ur Nanotechnologie, Karlsruhe Institute of Technology,
 76021 Karlsruhe, Germany}
 \affiliation{L.\ D.\ Landau Institute for Theoretical Physics RAS, 119334 Moscow, Russia}

\author{A.L. Burin}
\affiliation{Department of Chemistry, Tulane University, New Orleans, LA 70118, USA}

\author{I.V. Gornyi}
\affiliation{
 Institut f\"ur Nanotechnologie, Karlsruhe Institute of Technology,
 76021 Karlsruhe, Germany}
 \affiliation{A.\ F.\ Ioffe Physico-Technical Institute, 194021 St. Petersburg, Russia}
 \affiliation{L.\ D.\ Landau Institute for Theoretical Physics RAS, 119334 Moscow, Russia}

\author{R.A.Santos}
\affiliation{Department of Physics, Bar Ilan University, Ramat Gan 52900,
Israel }
\affiliation{Department of Condensed Matter Physics, Weizmann Institute of Science, Rehovot 76100, Israel}

\author{A.\ D.\ Mirlin}

\affiliation{
 Institut f\"ur Nanotechnologie, Karlsruhe Institute of Technology,
 76021 Karlsruhe, Germany}
\affiliation{\mbox{
 Institut f\"ur Theorie der Kondensierten Materie,
 Karlsruhe Institute of Technology, 76128 Karlsruhe, Germany}}
\affiliation{Petersburg Nuclear Physics Institute,  188300 St.~Petersburg, Russia}
\affiliation{L.\ D.\ Landau Institute for Theoretical Physics RAS, 119334 Moscow, Russia}

\begin{abstract}
We study the heat conductivity in Anderson insulators  in the presence of power-law interaction.
Particle-hole excitations built on localized electron states are viewed as two-level systems randomly  distributed in space and energy and coupled due to electron-electron interaction. 
A small fraction of these states form resonant pairs that in turn build a complex network allowing for energy propagation.  
We identify  the character of energy transport through this network  and evaluate  the thermal conductivity.  For physically relevant cases of 2D and 3D spin systems with $1/r^3$ dipole-dipole interaction (originating from the conventional $1/r$ Coulomb interaction between electrons), the found thermal conductivity $\kappa$ scales with temperature as  $\kappa\propto T^3 $ and $\kappa\propto T^{4/3}$, respectively. Our results may be of relevance also to other realizations of random spin Hamiltonians with long-range interactions.

\end{abstract}


\maketitle

\section{Introduction}
\label{section-intro}

 Wiedemann-Franz law (WFL) \cite{WF}  establishes a universal relation between the electric and thermal conductivity of metals. 
 WFL is strictly valid in a model of non-interacting electrons elastically scattered by impurities \cite{Abrikosov}.
 Inelastic scattering as well as quantum corrections due to interplay of  interaction and disorder 
 \cite{Savona, Livanov, Catelani,Catelani07,Schwiete2015,Karen,GGM} violate the WFL but the deviations are usually small.
 
 In this paper we address the problem of thermal conductivity in Anderson insulator. 
Anderson localization \cite{Anderson1958} drives electrons  
in a disordered metal  into an insulating phase, thus strongly  suppressing  the electric transport at low temperature $T$. In this situation transport is typically of variable-range hopping 
or activated nature so that the  suppression  of electrical conductivity is exponential. 
We show below that, in the presence of a power-law interaction, the thermal conductivity is a power-law function of $T$, so that the WFL is very strongly violated. We calculate $\kappa$ for arbitrary spatial dimensionality and for arbitrary exponent of the interaction law. Our results for physical situation of $1/r$ Coulomb interaction in two-dimensional (2D) and three-dimensional (3D) systems should be of experimental relevance.
 
One of motivations for our work was provided by experiments on 2D systems in the regime of integer and fractional quantum Hall effect (QHE). The Anderson localization of charged bulk excitations plays a crucial role in the outstanding accuracy of QHE quantization. In other words, these systems are perfect insulators with respect to electric current. On the other hand, recent measurements of energy transport via  quantum dot  \cite{Altimiras,Yacoby} and shot noise \cite{Inoue} thermometry,  detected  a considerable thermal conductivity  flowing through the bulk of fractional QHE with various filling fractions $\nu$ \cite{Altimiras,Inoue} and of $\nu=1$ integer QHE \cite{Yacoby}.  Since the bulk thermal transport was absent at filling fractions $\nu=2,\ 3$, it was definitely a property of the electronic system. 
Related results on a mysterious leakage of energy from the edge at $\nu=1$ were obtained in Ref.~\onlinecite{Eisenstein}.
These findings  prompted us to explore a state of the electronic system which is a charge insulator and an energy conductor at the same time. 
  
We consider a system deep in the Anderson insulator phase, with all single particle states being localized.
Electron hops on distances considerably exceeding the localization length $\xi$ are exponentially suppressed and will be discarded in our consideration. 
 In this approximation, the full Hilbert space of particle-hole excitations reduces to a subspace built by small (of size $\sim \xi$) two-level systems which can be represented in terms of spin 1/2 operators $S_i$. As all other relevant spatial scales will be much larger than $\xi$ at sufficiently low temperatures, one can view these spins as point-like objects.  The electron-electron interaction leads to an interaction between the spins, leading to an effective Hamiltonian  of the form \cite{Burin1998,Levitov,Burin2006,Demler}
\begin{eqnarray}
H=\sum_i\epsilon_i S_i^z + \sum_{ij}\frac{t_{ij}}{r_{ij}^\alpha}(S_i^+S_j^-+h.c.)+
\frac{V_{ij}}{{r_{ij}^\beta}}S_i^zS_j^z\,.
\label{Hamiltonian}
\end{eqnarray}
This spin model is  characterized by exactly zero electrical conductivity while the energy transport may still be finite.

If one starts from the conventional $1/r$  Coulomb interaction, the particle-hole pairs separated by a large distance  ($\gg \xi$)
exhibit a dipole-dipole interaction with $\alpha=\beta=3$. It is instructive to consider, following Ref.~\onlinecite{Demler}, a more 
general case, allowing  for  arbitrary  
\be
\alpha \geq \beta > d. 
\label{conditions-alpha-beta}
\ee
For $d>2$ our results will be also applicable in the borderline situation $\alpha \geq \beta = d$, including the important case of dipole-dipole interaction in 3D systems. 
We assume spin positions $i$ with uniform spatial distribution with density $\rho$ and the random Zeeman fields (splittings of two-level systems) $\epsilon_i$ uniformly distributed over the energy interval $(-W/2,W/2)$. The matrix elements prefactors 
$t_{ij}$ and  $V_{ij}$ are in general random  (in particular, have a random sign) with characteristic values $t_{ij} \sim t$, and $V_{ij} \sim V$. For $\alpha=\beta$ we will assume that $V\gtrsim t$. 

While the assumption of uniform distribution of energies $\epsilon_i$ is natural in other realizations of the spin Hamiltonian (\ref{Hamiltonian}) (some of them will be discussed below in this Section), it requires a comment in the case when this Hamiltonian represents an effective description of an interacting fermionic system. Indeed, for free fermions the density of states of dipole (particle-hole) excitations on top of the ground state (filled Fermi sea) is not constant for small energies $\epsilon$ but rather is linear in $\epsilon$. This is an immediate consequence of the fact that, in order to have a particle-hole pair with an energy $\epsilon$, the particle and the hole should both have energies below $\epsilon$. However, the Coulomb interaction essentially modifies this result, leading to a constant density of states of dipole excitations at not too high energies \cite{SE}. This justifies our assumption of constant density of energies $\epsilon_i$ \cite{footnote-dipole-dos}. 

In addition to the electronic realization discussed above, the spin Hamiltonian (\ref{Hamiltonian}) arises in several other physical contexts. We briefly discuss some of them. 

An important realization of the Hamiltonian (\ref{Hamiltonian}) is provided by amorphous materials (glasses) which show remarkable peculiarities in thermal transport and specific heat \cite{Zeller71,Hunklinger86}. To explain these anomalies, a model of two-level systems---atoms or groups of atoms that can tunnel between two nearly degenerate configurations---was proposed in Refs.~\onlinecite{Anderson72,Phillips72}.  Later work emphasized importance of interactions between the two-level systems \cite{Yu88,Burin89,Burin1998,Burin94,Classen00}. Recent experiments with superconducting circuits \cite{Lisenfeld10,Burnett14} provided a direct way to monitor the two-level systems and demonstrated a crucial role of interactions between them. We refer the reader to Refs.~\onlinecite{Burin94,Burin04} for applications of the model (\ref{Hamiltonian}) (in spatial dimensionality $d=3$ and with dipole interaction, $\alpha=\beta=3$) to the analysis of relaxation in glasses at low temperatures. 

Further, as discussed in Refs.~\onlinecite{Barnett06,Gorshkov11,Demler},  the Hamiltonian (\ref{Hamiltonian}) arises as a description of an ensemble of dipolar molecules in an optical lattice or of spin defects in a solid-state system. An experimental implementation of the dipolar-molecule setup was reported in Refs.~\onlinecite{Yan13,Hazzard14}. 
Experimental realization of a one-dimensional system of trapped ions with tunable long-range interaction \cite{Porras04} that can be approximated by a power law with a tunable exponent has been reported in Refs.~\onlinecite{Monroe13,Monroe15}. 

The structure of this article is as follows. In Sec. \ref{section-Resonant-spin-networks}, following Refs. \onlinecite{Levitov, Burin1998, Burin2006, Demler}, we identify the basic low-energy delocalized 
degrees of freedom (``networks'') in the Hamiltonian (\ref{Hamiltonian})  for various values of the exponents $\alpha$ and $\beta$. We find the criterion for the phase transition into the many-body localized phase and summarize  the phase diagram of the system.  Sections \ref{section-Thermal-transport-Optimal-network}, \ref{section-Thermal-transport-High-energy-excitations},  \ref{section-Thermal-transport-Low-energy-excitations} and \ref{section-tails}  are devoted to the quantitative analysis of the heat conductance in the parameter regime $d>\alpha\beta/(\alpha+\beta)$  within the approximation that neglects the ``spectral diffusion'' phenomenon \cite{Klauder62,Galperin83,Burin94} discussed later in Sec. \ref{section-Spin-relaxation-and-spectral-diffusion}.  They deal
with the thermal transport by  the ``optimal''  low-energy degrees of freedom (Sec. \ref{section-Thermal-transport-Optimal-network}), optimal-network-assisted transport by high-energy excitations (Sec. \ref{section-Thermal-transport-High-energy-excitations}) as well as with transport due to ultra-low-energy networks (Sec.\ref{section-Thermal-transport-Low-energy-excitations}) and the power-law tails in the hopping (Sec. \ref{section-tails}). Among all those mechanisms for heat conductance we find the optimal-network-assisted transport  to dominate in most cases.  
In section Sec. \ref{section-Pseudo2-spin-networks-and-thermal-transport} we deal with the thermal transport in the situation when the parameters of the model satisfy $\beta/2<d<\alpha\beta/(\alpha+\beta)$. This parameter range (which does not exist for $\alpha=\beta$ and, in particular, in the physically most interesting case of dipole-dipole interaction, $\alpha=\beta=3$) requires the analysis of a more complicated network as compared to more conventional situation
$d>\alpha\beta/(\alpha+\beta)$ explored in Secs.~\ref{section-Thermal-transport-Optimal-network}, \ref{section-Thermal-transport-High-energy-excitations},  \ref{section-Thermal-transport-Low-energy-excitations}  and \ref{section-tails}.
In Sec. \ref{section-Localization-threshold} we discuss the scaling  of the localization threshold for the many-body states with the system size and include in consideration the spectral diffusion  whose implications for the thermal transport are analyzed in Sec. \ref{section-Spin-relaxation-and-spectral-diffusion}.
We close the paper by summarizing our results  in Sec. \ref{section-summary}.

 \section{Resonant spin networks}
\label{section-Resonant-spin-networks}

In this section we summarize  the phase diagram of the system described by the Hamiltonian (\ref{Hamiltonian}) as derived in Refs. \onlinecite{Levitov, Burin1998, Burin2006, Demler}. We identify the delocalized low-energy excitations in the system (that exist for $\beta<2d$) and establish the corresponding effective theory.  We focus on the case of low temperature and strong disorder, assuming
that $T \ll W$ and $t\rho^{\alpha/d} \ll W$.

\subsection{Direct spin network}
\label{s21}

To analyze whether the system is in conducting or insulating phase,
one first performs a counting of resonant spins in spirit of Ref.~\onlinecite{Anderson1958}. 
Two spins $i$ and $j$ form  a resonant pair under the condition
\begin{equation}
\label{res}
 |\tilde{\epsilon}_i-\tilde{\epsilon}_j|\lesssim\frac{t}{R_{ij}^\alpha} \,,
\end{equation}
where   $R_{ij}=|r_i-r_j|$ is a distance between them and $\tilde{\epsilon}_{i}\equiv\epsilon_{i }+\sum_{k}V_{ik}S^z_i S^z_k/r_{ik}^\beta$
is the energy of the spin $i$   with the contribution of its interaction with the neighboring  spins taken into account. The interaction-induced correction to $\epsilon_{i }$ makes the definition of resonance  for two spins dependent on the state of other spins in the system. While this may influence the thermal transport  via   the ``spectral diffusion'' phenomenon,  it is of no importance for the power-counting arguments involved in the determination of the phase diagram of the system. 
Indeed, in the many-body localized phase the spins are essentially frozen. Thus, discussing the stability of the localized phase and the basic delocalization mechanisms we can safely ignore the fact that the definition of resonances depends on the state of the system.
 We will proceed in this way in Secs. \ref{section-Resonant-spin-networks} ---   \ref{section-Pseudo2-spin-networks-and-thermal-transport} (and drop tilde in $\tilde{\epsilon}_i$), returning  to the effect of spectral diffusion in Secs. \ref{section-Localization-threshold} and \ref{section-Spin-relaxation-and-spectral-diffusion}.

If one chooses a particular spin, the average number of its resonant partners within  
the  layer $R < |r_i-r_j| < 2R$ is given by
\begin{equation}
N_1(R)\sim \rho R^d\times \frac{t}{WR^\alpha} =\frac{t\rho }{W} R^{d-\alpha}.
\label{N1}
\end{equation}
Here the factor $\rho R^d$ is the total number of spins within the volume $R^d$ around the chosen one, while  the factor $t/WR^{\alpha}$ takes into account  the resonance condition. When the average is small, $N_1(R) \ll 1$, it has a meaning of the probability for a spin to have a resonant partner within the above layer. 

For  $d >\alpha$ the number $N_1(R)$ grows with $R$ and an infinite network of resonating spins emerges  so that the system is in the delocalized phase.
On the other hand, for  $d<\alpha$ the number $N_1(R)$ decreases with $R$. In this case the spin is in average in resonance only with a finite number of spins,
\begin{equation}
N_{\rm total} \sim \int_{\rho^{-1/d}}^\infty\frac{dR}{R}  N_1(R)
=\frac{t \rho^{\alpha/d}}{W}.
\end{equation}
In the assumed regime of strong disorder, the average number of resonant partners (or, equivalently, the probability for a given spin to participate in a resonant pair) satisfies  $N_{\rm total} \ll 1$. 
For a non-interacting  problem this implies Anderson localization \cite{Anderson1958}. 
The dimension $d=\alpha$ is critical  in the non-interacting case. The corresponding model was previously studied in a number of works
\cite{Anderson1958,Levitov90,Levitov,Mirlin_Evers} and is known to show a fractal behavior, in particular, an
anomalous diffusion with time dependence of the typical displacement of the form $r\sim t^{1/d}$. 

Below we focus on  the case $d<\alpha $  (and $N_{\rm total}\ll 1$), in which a non-interacting system is in the insulating phase.
It turns out, however, that the counting argument presented above is insufficient for the spin problem (which corresponds to an interacting fermion problem).
As was pointed out in Refs.~\onlinecite{Burin89,Burin1994,Burin98,Burin2006},  there is a more efficient way of building a
connected spin network   and $d=\alpha$ is not a critical dimension in the interacting problem. We give an account of the procedure in Secs. \ref{section-Resonant-spin-networks2} and \ref{section-Resonant-spin-networks3}.

\subsection{Pseudo-spin network}
\label{section-Resonant-spin-networks2}

To proceed further  one first  selects  pairs of spins $i,j$  that  are at  resonance. Each such resonant pair has four energy levels. One considers two of them that correspond to $S_i^z+S_j^z=0$; their energy splitting is  $\sim t/R_{ij}^\alpha$. This two-level system can be described by a new pseudo-spin $\tau_a$ with spatial size $R_{ij}$ and energy $E_{ij} \sim t/R_{ij}^\alpha$. 
To participate in the dynamics at a given temperature $T$, the pseudospin should be built out of spins with 
energies within the temperature stripe, $|\epsilon_i|, |\epsilon_j| \lesssim T$.  While two high-energy spins $|\epsilon_i|, |\epsilon_j | \gg T$ can also form a resonant pair with a small splitting, $E_{ij} \lesssim T$, thermal occupation numbers of both states of such a pseudospin will be exponentially suppressed. Such exponentially small contributions are of no relevance for our consideration and will be neglected. Below we only consider the pseudospins that are ``active'' at a temperature $T$.

\begin{figure}
\includegraphics[width=230pt]{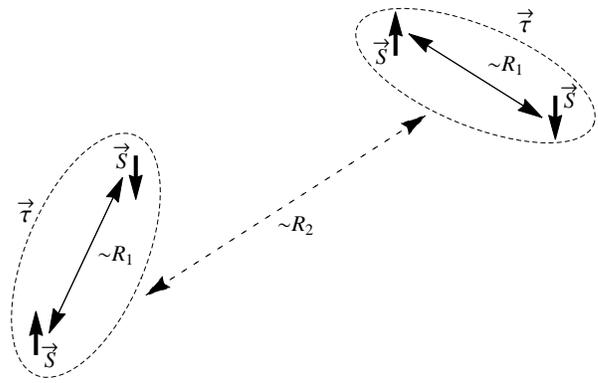}
\caption{ Formation of a resonant network of pseudo-spins (Sec. \ref{section-Resonant-spin-networks2}). Each pseudo-spin $\tau$ consists of two (resonating) spins $\vec{S}$ at distance $R_1$. Two pseudo-spins at distance $R_2$ are at resonance provided that the condition (\ref{condition-resonance}) is fulfilled. To construct the pseudo$^2$-spin network, this construction is iterated once more (Sec. \ref{section-Resonant-spin-networks3}).}
\label{net}
\end{figure}

After  pseudo-spins are constructed, one analyzes their connectivity,  see Fig. \ref{net}.
 The interaction between the pseudo-spins is provided by the last term in the Hamiltonian (\ref{Hamiltonian}). A pair of pseudo-spins of a spatial size $R_1$ is at resonance if
\begin{equation}
\frac{V}{R_2^\beta}\gtrsim |E_{12}-E_{34}|\sim \frac{t}{R_1^\alpha}\,.
\label{condition-resonance}
\end{equation} 
The condition (\ref{condition-resonance}) determines a spatial distance $R_2^*$ within which a pseudo-spin of the size $R_1$ is at resonance with any other pseudo-spin of the same size,
\begin{equation}
R_2^*=\left(\frac{V}{t}R_1^\alpha\right)^{1/\beta}.
\label{R2star}
\end{equation}
Since  $\beta \leq \alpha$ (and $V\gtrsim t$ for $\alpha=\beta$), the distance between  pseudo-spins is larger than (or of the order of) their size, $R_2^* \gtrsim R_1$.  Next, one  choses a pseudo-spin of the size $R_1$ and 
count the pseudo-spins  that are at resonance  with it  inside the shell $R_2 < R <2 R_2$. 
We denote this number  $N_2(R_1,R_2)$, adopting notations of Ref.~\onlinecite{Demler}.
The total number of (``active'') pseudo-spins of size $R_1$ in this volume is 
 $\sim (T/W) \rho  R_2^dN_1(R_1) $. 
Though the spatial density of pseudo-spins $\rho_{\rm ps}$ is lower than the density $\rho$ of the original spins, 
\begin{equation}
 \rho_{\rm ps}(R_1)\sim\rho N_1(R_1) T/W \sim t\rho^2 T R_{1}^{d-\alpha}/W^2\,,
 \label{ps-density}
\end{equation}  
the pseudospins
 are distributed over a  narrow  energy interval $t/R_1^\alpha$. For $V/R_2^\beta \gtrsim t/R_1^\alpha$ all $R_1$ pseudo-spins in the considered volume are at resonance; in the opposite case, the  fraction of pseudo-spins that are at resonance with a given one is $(V/R_2^\beta)/(t/R_1^\alpha)$.  This yields 
\begin{equation}
N_2(R_1,R_2)= \rho_{\rm ps}(R_1) R_2^d \: {\rm min}\bigg\{1,\frac{V/R_2^\beta}{t/R_1^\alpha}\bigg\} .
\label{N2}
\end{equation}

\begin{figure}[ht]
\vspace{0cm}
\includegraphics[width=0.4\textwidth]{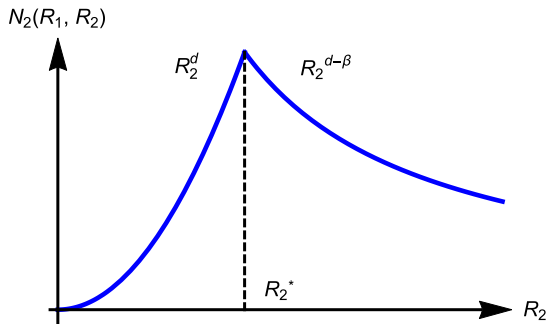}
\caption{Number of pseudo-spins of size $R_1$ in the shell $R_2<r<2R_2$  that are in resonance with a given pseudo-spin of size $R_1$.
\small}
\label{fig1}
\end{figure}

For $\beta<d$ the number of resonant pseudo-spins $N_2(R_1, R_2)$, Eq. (\ref{N2}), diverges as $R_2\rightarrow\infty$ at fixed $R_1$. Thus, in full analogy with Sec. \ref{s21}, the pseudo-spins 
with any given $R_1$ form infinite resonant networks that support delocalized excitations. 

The delocalization occurs in a slightly more complicated fashion for $\beta>d$, the case of our primary interest. As illustrated  in Fig.~\ref{fig1},  the number of resonant pseudo-spins $N_2(R_1,R_2)$ has in this situation  a sharp maximum as a function of $R_2$ at $R_2 \simeq R_2^*$.  At the optimal distance $R_2^*$, the typical number of resonant pseudo-spins  of the size $R_1$  is
\begin{equation}
\label{N2*}
N_2(R_1,R_2^*(R_1))\sim \frac{t\rho^2 T}{W^2}\left(\frac{V}{t}\right)^{d/\beta}R_1^{\frac{d\alpha}{\beta}+d-\alpha}\,.
\end{equation}
For 
\be
d > \alpha\beta/(\alpha+\beta)
\label{condition-deloc}
\ee
 this  function monotonously increases with increasing $R_1$, and the system 
is in the energy-conducting phase. Therefore, Eq.~(\ref{condition-deloc}) is the condition for the delocalization of the pseudospin network.
For $\alpha=\beta$ this condition  reduces to $d>\alpha/2$, which is the result obtained in Ref.~\onlinecite{Burin2006}. The generalization (\ref{condition-deloc}) was found in Ref.~\onlinecite{Demler}. 

If $d < \alpha\beta/(\alpha+\beta)$, the pseudo-spins fail to form a connected network (at least, at low enough temperatures). However, it does not yet mean that the system is necessarily  an insulator, as the next step 
of the hierarchical construction reveals.

\subsection{Pseudo$^2$-spin network}
\label{section-Resonant-spin-networks3}

In order to explore the possibility of delocalization for  $d<\alpha\beta/(\alpha+\beta)$, we follow Ref.~\onlinecite{Demler} and go to the next level of hierarchy, proceeding once more in  spirit of Secs. \ref{s21} and \ref{section-Resonant-spin-networks2}. 
We identify (rare) resonant pairs of pseudo-spins and replace them by the effective pseudo$^2$-spins $\sigma$. Each pseudo$^2$-spin $\sigma$ is an object of some size $R_2$ consisting of two pseudo-spins of size $R_1$ satisfying the resonance condition~(\ref{condition-resonance}). 

The spatial density of pseudo$^2$-spins is 
\begin{equation}
 \rho_{{\rm p}^2{\rm s}}\sim \rho_{\rm ps}(R_1)N_2(R_1, R_2)\sim (T/W) \rho N_1(R_1)N_2(R_1, R_2),
 \label{p2s-density}
\end{equation}
and their energy distribution has a width $\sim V/R_2^\beta$. Further, two pseudo$^2$-spins at a distance $R_3$ have a typical interaction energy $\sim V/R_3^\beta$. 

We estimate now the number of pseudo$^2$-spins at resonance with a given pseudo$^2$-spin $\sigma$ of size $R_2$  (built out of spins of size $R_1$) and at a distance $R_3\gtrsim R_2$ from it. In analogy with Eq.~(\ref{N2}), we get
\begin{eqnarray}
 && N_3(R_1, R_2, R_3)\sim\rho_{{\rm p}^2{\rm s}} R_3^d\times\frac{V/R_3^\beta}{V/R_2^\beta}    \nonumber\\
 && \sim\frac{t^2 \rho^4 T^2}{W^4}R_1^{2d-2\alpha}R_2^d\:{\rm min}\!\left\{1, \frac{V/R_2^\beta}{t/R_1^\alpha}\right\}R_3^{d-\beta}R_2^\beta.
 \hspace{1cm}
\label{N3}
 \end{eqnarray}
 
 Clearly, the analysis of the pseudo$^2$-spin-network is of particular interest in the case $d<\alpha\beta/(\alpha+\beta)$, when the pseudo-spins do not form by themselves a resonant network. 
 As will be discussed below, this analysis  in fact makes sense already under weaker conditions,  $d<\beta$ and $d<\alpha/2$. (We recall that we always assume $\alpha\geq\beta$.)    Using the  inequalities $d<\beta$ and $d<\alpha/2$, we find that $N_3(R_1, R_2, R_3)$ is a monotonously decaying function of $R_1$ and $R_3$ attaining its maximum value
 \begin{equation}
  N_3(R_{1, \rm th}, R_2, R_2)\sim\frac{tV\rho^4T^2}{W^4}R_{1, th}^{2d-\alpha}R_2^{2d-\beta}
  \label{N3*}
 \end{equation}
 at $R_3=R_2$  and $R_1=R_{1,\rm th}(T)$, with
 \begin{equation}
R_{1, \rm th} (T)\sim (T/t)^{-1/\alpha}
\label{R-thermal}
\end{equation}
 being the thermal pseudo-spin size determined by $t/R_{1, \rm th}^\alpha =T$. 
 
Equation (\ref{N3*}) implies that the pseudo$^2$-spins form a connected resonant network, and thus the system is in the delocalized state   for 
\begin{equation}
\label{p2s-deloc-condition}
d > \beta/2. 
\end{equation}
Under the opposite condition, $d<\beta/2$,  the system is an insulator,  at least on the level of pseudo$^2$-spins. 

 The critical dimensionality $d = \beta/2$ was found in Refs.~\onlinecite{Burin89,Burin94,Burin98} in a model of spins in a randomly directed Zeeman field and with long-range Ising couplings (exponent $\beta$). In this model, the dynamics occur only due to local fields, which places it into the same ``universality class''  as Eq.~(\ref{Hamiltonian}) with $\alpha=\infty$. In Ref.~\onlinecite{Burin2006} the result $d = \beta/2$ for the critical dimensionality was derived for the model (\ref{Hamiltonian}) with $\alpha = \beta$.  More recently, Ref.~\onlinecite{Demler} obtained the critical dimensionality $d = \beta/2$ for the model (\ref{Hamiltonian}) with arbitrary  $\beta \le \alpha$.
 
In order to study the regime $\beta>2d$ one may be tempted to further iterate the construction. It was, however, argued in  Refs.~\onlinecite{Burin89,Demler} that this does not lead to any further reduction of the critical dimensionality. Thus, the line $d=\beta/2$ marks the true phase transition to the (low-temperature) many-body localized phase. 

\begin{figure}[ht]
\includegraphics[width=230pt]{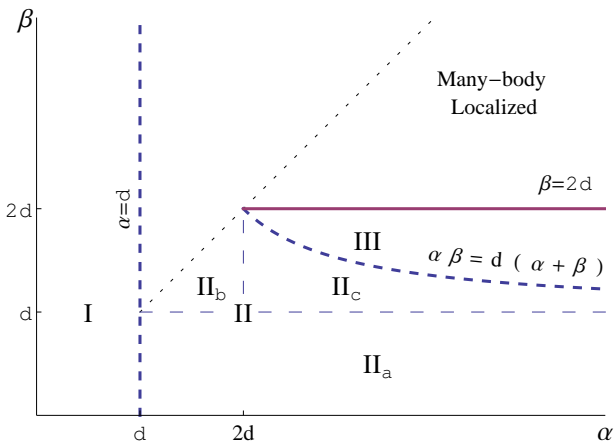}
\caption{\small
Phase diagram of the system. We concentrate on the parameter range $\alpha \ge \beta$, which is the part of the phase diagram limited by the dotted line.  The full 
line $\beta=2d$ designates the phase transition to the many-body localized phase.
Thick dashed lines divide the delocalized phase into the regions I, II, and III that are distinguished by the basic delocalized low-energy degrees of freedom. Specifically, the lowest-level resonant network is that of spins in the region I, of pseudo-spins in the region II, and of pseudo$^2$-spins in the region III. Thin dashed lines subdivide the region II into regions with different structure of the optimal pseudo-spin and/or pseudo$^2$-spin networks.  In the region II$_{\rm a}$, the pseudo-spins of any size become delocalized by themselves, in full analogy with delocalization of spins in region I. In  the parts II$_{\rm b}$ and II$_{\rm c}$ the delocalization mechanism is much more intricate, and only pseudo-spins of sufficiently large size $R_1$ become connected. In the region II$_{\rm c}$ delocalized pseudo-spin and pseudo$^2$-spin networks coexist.
In this paper, we focus on regions II$_{\rm b}$, II$_{\rm c}$, and III.
}
\label{Fig:PhaseDiagram}
\end{figure}

The phase diagram of the system is summarized in Fig. \ref{Fig:PhaseDiagram}. We concentrate on the parameter range $\alpha \ge \beta$, i.e., the part of the phase  diagram limited by the dotted line.  The full line $\beta=2d$ designates the phase transition to the many-body localized phase. The thick dashed lines divide the delocalized phase into three regions I, II, and III that differ in the basic delocalized low-energy degrees of freedom. Specifically, in the region I, a connected resonant spin network can be built. In the region II, spins do not form a connected network but pseudo-spins do form it. Similarly, in the region III, neither spins nor pseudo-spins form a connected network, but it can be built out of pseudo$^2$-spins. 

The region II is further subdivided by thin dashed lines into several parts.   As was already mentioned in Sec. \ref{section-Resonant-spin-networks2}, the line $\beta=d$ separates the regions  with different structure of the connected pseudo-spin networks. In the region II$_{\rm a}$ (with $\beta<d$), the pseudo-spins of any size become delocalized by themselves, in full analogy with delocalization of spins in region I. 
On the other hand, in  the parts II$_{\rm b}$ and II$_{\rm c}$ of the region II where the delocalization mechanism is much more intricate, and only pseudo-spins of sufficiently large size $R_1$ become connected. In this paper, we focus on regions II$_{\rm b}$, II$_{\rm c}$, and III.

The construction of the connected pseudo$^2$-spin network is particularly clear in the region III of the phase diagram where no connected pseudo-spin network exists. It turns out, however, that the pseudo$^2$-spin network is also  meaningful in the region II$_{\rm c}$ of the phase diagram, despite the existence of a connected pseudo-spin network. The reason for this is as follows.  The conducting pseudo-spin network is built of pseudo-spins of large size $R_1$ such that $N_2(R_1,R_2^*(R_1)) \gtrsim 1$. On the other hand, the conducting pseudo$^2$-spin network is built out of pseudo-spins with a much smaller size $R_1\sim R_{1, \rm th}(T)$ for which $N_2(R_1, R_2^*(R_1))\ll 1$. Such small (or, equivalently, high-energy) pseudo-spins do not form by themselves a resonant network. 
Thus, in the region  II$_{\rm c}$ the pseudo-spin and pseudo$^2$-spin networks coexist, as they are formed by pseudo-spins with parametrically different sizes (or energies).  

The situation is different in the region  II$_{\rm b}$, where the function $N_3(R_1, R_2, R_3)$ achieves its maximum at $R_3\sim R_2\sim R_2^*(R_1)$. Thus, in this region, the connected networks of pseudo$^2$-spins would be just the same as the connected pseudo-spin networks.   Therefore, the extension of the hierarchical construction to the pseudo$^2$-spin level does not bring anything new in the region  II$_{\rm b}$, and one should stop at the pseudo-spin level. 

Numerical analysis of the one-dimensional (1D) problem (\ref{Hamiltonian}) in Ref.~\onlinecite{Demler} with $\alpha=\beta$ supported the existence of a transition at a critical value of the exponent satisfying $1 < \alpha_c < 3$, consistent with the analytical prediction $\alpha_c=2$.
In Ref.\onlinecite{Pino} the 1D model (\ref{Hamiltonian}) was numerically studied for $\alpha=\beta=3$ and 5, and the many-body localization was confirmed. A subsequent detailed study in Ref.~\onlinecite{Burin15b} provided a further numerical evidence in favor of $\alpha_c=2$.

The phase diagram of  Fig. \ref{Fig:PhaseDiagram} will guide our discussion of thermal transport in the rest of the paper. 
In Secs. \ref{section-Thermal-transport-Optimal-network}, \ref{section-Thermal-transport-High-energy-excitations} and \ref{section-Thermal-transport-Low-energy-excitations} we focus on the region II of the phase diagram (more precisely, in subregions II$_{\rm b}$ and II$_{\rm c}$) and analyze the heat  conductivity  due to a connected pseudo-spin network. In Sec. \ref{section-Pseudo2-spin-networks-and-thermal-transport} we investigate the thermal transport caused by the connected network of pseudo$^2$-spins, which is of primary importance in the region III of the parameter space. We then also discuss the influence of pseudo$^2$-spins on the transport in the  region and II$_{\rm c}$.   

In Secs. \ref{section-Thermal-transport-Optimal-network} --- \ref{section-Pseudo2-spin-networks-and-thermal-transport} we will consider pseudo-spins  and pseudo$^2$-spins as rigid entities and discard other degrees of freedom. Effects of spectral diffusion which leads to modification of resonant spin pairs forming pseudo-spins [see discussion after Eq. (\ref{res})] are considered in Sec. \ref{section-Spin-relaxation-and-spectral-diffusion}.

It is worth emphasizing at this point that we aim at the description of the the energy transport at the delocalized phase, $d > \beta/2$. We will not study properties of the system at the critical dimensionality $d = \beta/2$. It is expected that observables show at criticality a multifractal behavior requiring a careful analysis of full distribution functions, as has been performed for non-interacting problems in Refs.~\onlinecite{Levitov90,Levitov,Mirlin_Evers}. We leave the investigation of this critical regime as an interesting prospective for future research. On the other hand, away from criticality (on the delocalized side), conducting networks are formed that lead to conventional diffusive energy transport. Typical parameters of these networks and resulting transport characteristics can be obtained by scaling analysis presented in this paper. Of course, local characteristics (such as, e.g., a distance between two nearest neighbor pseudospins in a network) do fluctuate but the magnitude of these fluctuations is of order of the typical value. Such fluctuations are expected to influence only numerical prefactors  in expressions for transport characteristics. We do not try to evaluate such prefactors in this paper and omit them in the formulas below. We keep, however, the dependence on all physical parameters of the problem. 

\section{Thermal transport: Optimal network}
\label{section-Thermal-transport-Optimal-network}

We begin our study of thermal transport with the analysis of thermal conductivity originating from pseudo-spin networks in the parameter range $\alpha\beta/(\alpha+\beta) < d < \beta$ (regions II$_{\rm b}$ and II$_{\rm c}$ of the phase diagram in Fig. \ref{Fig:PhaseDiagram}).
To study  thermal transport,  we  model the system by a set of pseudo-spin  networks.
Each network consist  of a pseudo-spins of approximately  same size (within factor of two). We will first discard coupling between the network and will discuss its role later on.
The Hamiltonian of a network of pseudo-spins  of a size $\sim R_1$ is given by\cite{Demler}
\begin{equation}
H=\sum_i\left( E_i^x\tau_i^x+E_i^z\tau_i^z\right)+\sum_{i,j}u_{ij}\tau_i^z\tau_j^z\,.
\label{H-pseudospins}
\end{equation}   
The energies of pseudo-spins are randomly distributed over the  band of the width $t/R_1^\alpha$,
\be
E_i^x \sim E_i^z \sim t/R_1^\alpha,
\label{ps-energy}
\ee
and their density in space is given by Eq. (\ref{ps-density}).

The interaction between the pseudo-spins is random and has the magnitude $u_{ij}\sim V / |r_i-r_j|^\beta$.
As explained in Sec.~\ref{section-Resonant-spin-networks}, the dominant role is played by interactions at the scale $\sim R_2^*(R_1)$ where $N_2(R_1,R_2)$ has a maximum as a function of $R_2$. We can thus approximate the
pseudo-spin interaction by 
\begin{equation}
\label{multistep-distribution}
u_{ij} \sim \left\{
\begin{array}{ll}
\displaystyle \frac{V}{[R_2^*(R_1)]^\beta} \sim \frac{t}{R_1^\alpha}\,,  \qquad & |r_i-r_j|<R_2^*(R_1) ; \\
\\
0 \,, & {\rm otherwise.}
\end{array}
\right.
\end{equation}
While the neglected terms with $|r_i-r_j| \gg R_2^*(R_1)$  are of no importance for delocalization
(since they decay as $|r_i-r_j|^{-\beta}$ with $\beta>d$), one still should check what is their contribution to the transport. Indeed, it is known that a power-law hopping may lead to superdiffusive behavior (Levy flights). We will return to this question in Sec.~\ref{section-tails} and show that the superdiffusive behavior does not arise (in the considered range of $\alpha$ and $\beta$) in the cases of our main interest $d = 2$, 3 but may be important for 1D systems.

\begin{figure}[ht]
\vspace{0cm}
\includegraphics[width=0.3\textwidth]{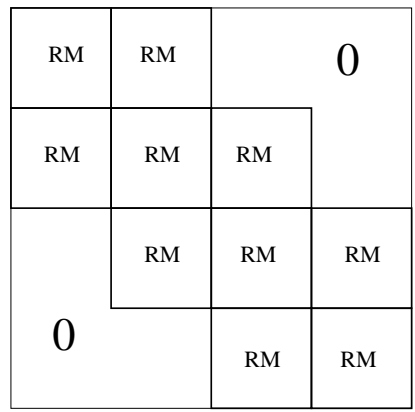}
\vspace{0.2cm}
\caption{Schematic representation of the random banded matrix approximation to the Hamiltonian of the network of pseudo-spins of a given size $R_1$. Each block has a size $R_2^*(R_1)$. Pseudo-spins within each block and between adjacent blocks are strongly coupled by interaction. Couplings at longer distances are neglected. 
}
\label{fig2}
\end{figure}
\vspace{0cm}

It is convenient to perform  a unitary  transformation such that the``Zeeman field'' for each pseudo-spin points in $z$ direction.   
The Hamiltonian takes then the form 
\begin{equation}
\label{rotated}
H=\frac{1}{2}\sum_i\epsilon_i\tilde{\tau}^z_i+\frac{1}{2}\sum_{i,j} u_{ij} (\vec{n}_i\vec{\tilde{\tau}}_i)(\vec{n}_j\vec{\tilde{\tau}}_j) \,,
\end{equation}
with random energies $\epsilon_i\sim \sqrt{(E_i^x)^2 + (E_i^z)^2} \sim t/R_1^\alpha$, random interactions $u_{ij}$, Eq.~(\ref{multistep-distribution}), and random unit vectors $\vec{n}_i$ in $x-z$ plane.
 
By virtue of Jordan-Wigner transformation, the spin problem can be mapped onto that of interacting fermions. Let us first neglect the fermion interaction; we will return to it below. We get then a random matrix problem with the Hamiltonian belonging to a $d$-dimensional version of a random banded matrix  ensemble, see Fig.~\ref{fig2}.  Each  box in this figure represents a random matrix of size
$N_2[R_1,R_2^*(R_1)]$;  it contains matrix elements between sites $i$ located within the volume  $R_2^*(R_1)$. 
All diagonal entries are $\epsilon_i\sim  t/R_1^\alpha$. Non-diagonal entries within each box as well as matrix elements between adjacent boxes are random hopping amplitudes $u_{ij}\sim  t/R_1^\alpha$. All other matrix elements have been neglected.

The key dimensionless parameter characterizing the connectivity of the network is $N_2[R_1,R_2^*(R_1)]$, which is nothing but the dimensionless conductance at the ``ultraviolet'' scale $R_2^*(R_1)$. If this number is large, $N_2[R_1,R_2^*(R_1)]\gg 1$, the network is conducting (``connected''); if it is small, $N_2[R_1,R_2^*(R_1)]\ll 1$, the network is in the localized regime (``disconnected''). These two regimes are separated by a critical value $N_2[R_1,R_2^*(R_1)]\sim 1$, which corresponds, according to Eq.~(\ref{N2}), to the following size of pseudo-spins:
\begin{equation}
\label{R1}
R_1(T) \sim \bigg[\frac{Tt\rho^2}{W^2}\left(\frac{V}{t}\right)^{\frac{d}{\beta}}\bigg]^{-\frac{\beta}{(\alpha+\beta)d-\alpha\beta}}.
\end{equation}
As follows from Eq.~(\ref{N2}), the number  $N_2[R_1,R_2^*(R_1)]$ is a monotonously growing function of $R_1$. 
Thus networks  made of pseudo-spins of size larger (smaller) than $R_1(T)$ are connected (respectively, disconnected)  \cite{note-optimal-pseudospin-size}.

The above conclusion of the delocalization transition that takes place with increasing dimensionless conductance $N_2$ is obvious in  a three-dimensional system (or, more generally, for $d>2$). On the other   hand, it is less trivial for $d\le 2$, since  noninteracting system is always localized in these dimensionalities. (For $d=2$ this concern is, in fact, somewhat academic, since the localization length grows exponentially with $N_2$ and becomes larger than any realistic sample size for the conductance $N_2\gtrsim 5$.)  At this point one should recall, however, that our system is, in fact, interacting and
the dephasing length due to inelastic processes
is of the same order $\sim R_2^*(R_1)$ as the localization length for the network with $N_2\sim 1$. This is because the network with $N_2\sim 1$ does not have any small dimensionless parameter: all relevant energy scales are of the same order.  With $N_2$ increasing beyond unity, the localization length increases and becomes much larger than the dephasing length, which, according to common wisdom \cite{altshuler81}, ensures delocalization. Thus, the condition $N_2\sim 1$, or, equivalently, Eq.~(\ref{R1}) marks the transition from connected (delocalized) to disconnected (localized) networks, independently of the spatial dimensionality $d$. 

We will term a delocalized network with $R_1(T)$ given by Eq.~(\ref{R1}) (with numerical coefficient different by, say, factor of two from the critical value) the optimal network. We will show below that, under certain conditions on the exponents $\alpha$ and $\beta$, the thermal transport is dominated by this network.

The contribution to the thermal conductivity from a connected network can be estimated as follows: 
\begin{equation}
\kappa(R_1) \sim \frac{E^3(R_1)}{T^2}N_2[R_1,R_2^*(R_1)]  [R_2^*(R_1)]^{2-d}\,,
\label{kappa-r1}
\end{equation}
where $E(R_1)$ is the typical energy carried by excitations on the network with given pseudo-spin size $R_1$.
Two powers of excitation energy, $E^2(R_1)$, in Eq. (\ref{kappa-r1}) come from the energy vertices in the linear-response calculation. Further, an additional small factor of $E(R_1)/T$ reflects reduced sensitivity of the distribution function to temperature variations in the situation when the band width is much smaller than temperature, i.e., it originates from the product of the derivative of the Fermi function $\sim 1/T$ and the band width $\sim E(R_1)$.

For the optimal network we have $N_2\sim 1$ and Eq.~(\ref{kappa-r1}) reduces to
\begin{equation}
\kappa_* \sim \frac{E_*^3}{T^2}  [R_2^*(R_1(T))]^{2-d}\,,
\label{kappa-optimal}
\end{equation}
where 
\be
E_*=t/R_1^\alpha(T)
\label{E-star}
\ee
 is a typical energy of delocalized modes on the optimal network.
Let us emphasize that, in view of $d < 2\alpha\beta/(\alpha+\beta)$ [which follows from the conditions (\ref{conditions-alpha-beta})], this energy is parametrically smaller than thermal energy, $E_*  \ll T$.  According to Eq.~(\ref{kappa-optimal}), the heat  conductance of  the optimal network has the  temperature dependence
\begin{equation}
\kappa_* \propto T^{\mu_*}\,,
\label{kappa-scaling}
\end{equation}
with the exponent
\begin{equation}
\mu_* =\frac{5\alpha \beta -(2+d)\alpha-2d\beta}{(\alpha+\beta)d-\alpha\beta}\,.
\label{exp-mu}
\end{equation}
For the  physically relevant case of $d=2$ and  $\alpha=\beta=3$, we find $\mu_{*}=7$ and the thermal conductivity
\begin{equation}
\label{kappa-d2}
\kappa_* \sim \frac{t^6 V^6 \rho^{18}}{W^{18}}T^7\,.
\end{equation} 
In the case of $d=3$ and $\alpha=\beta=3$, the above calculation yields $\mu_{*}=4/3$ and
\begin{equation}
\label{kappa-d3}
\kappa_* \sim V^3 \left(\frac{t \rho^2}{W^2}\right)^{10/3}T^{4/3}\,.
\end{equation}

\begin{figure}
\includegraphics[width=230pt]{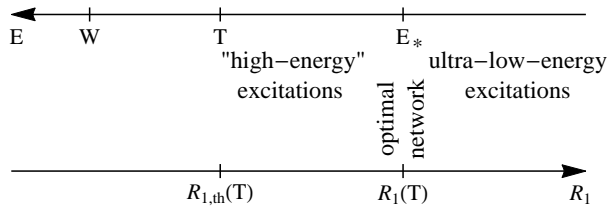}
\caption{  Characteristic energy scales ($W$ --- disorder, $T$ --- temperature, $E_*$ --- excitation energy in the optimal network) and corresponding pseudo-spin sizes. The optimal network is constructed and studied in Sec. \ref{section-Thermal-transport-Optimal-network}. Transport by ``high-energy'' excitations (which get delocalized only via the interaction with the optimal network) is explored in Sec. \ref{section-Thermal-transport-High-energy-excitations}. The ultra-low-energy excitations are discussed in Sec. \ref{section-Thermal-transport-Low-energy-excitations}. } 
\label{scales}
\end{figure}

This is, however, not the end of the story. What we have calculated by now is the contribution to $\kappa$ from the network that we called optimal. It remains to see, however, whether (and under what conditions) it is optimal indeed. This amounts to evaluating contributions of networks with pseudo-spin sizes $R_1$ much smaller and much larger than the optimal one, $R_1(T)$,  see Fig.~\ref{scales}  for the summary of energy and spatial scales in the problem. Furthermore, we remind the reader that the case $\alpha=\beta=d=3$ is on the borderline of the region of applicability of the theory, see the conditions (\ref{conditions-alpha-beta}). We should thus clarify whether the obtained results retain validity for this physically important case. 

\section{Thermal transport: High-energy excitations}
\label{section-Thermal-transport-High-energy-excitations}

In Sec.~\ref{section-Thermal-transport-Optimal-network} we considered separately networks corresponding to different pseudo-spin sizes $R_1$. We have shown that in this approximation there is a localization transition at a value $R_1(T)$ corresponding to an energy $E_*$:  excitations with smaller $R_1$ (or, equivalently, larger energies) are localized. However, in the full Hamiltonian there also contributions that couple pseudo-spins with different $R_1$. These terms have the same structure as given by the second term in Eq.~(\ref{rotated}), but now with pseudo-spins $\vec{\sigma}_i$ and $\vec{\sigma}_j$ having essentially different $R_1$ (and thus energies). This coupling, once taken in second (or higher) order of perturbation theory allows for decay processes of a high-energy excitation in two (or more) lower-energy excitations. These are real decay processes since the excitations with energy $\lesssim E_*$ are delocalized and thus form a continuous spectrum. Therefore, coupling to the low-energy excitations dephases excitations with higher 
energies and leads to their delocalization. Thus, excitation with energies higher than $E_*$ are also mobile and will contribute to the thermal transport. Clearly, pseudo-spins with small $R_1$ have low mobility: the interaction-induced decay rate of these ``nearly localized'' states is relatively long. On the other hand, the concentration of pseudo-spins with energies higher than $E_*$ is much larger than the concentration of pseudo-spins on the optimal network. Thus, it is not immediately clear which range of energies will give a dominant contribution to the transport.

\begin{figure}[ht]
\vspace{0cm}
\includegraphics[width=0.4\textwidth]{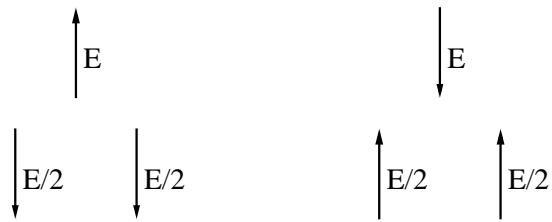} 
\vspace{0.2cm}
\caption{A process contributing to dephasing of high-energy pseudo-spins by those with lower energies. Here a spin with a high energy $E$ is dephased by a decay into two spins of roughly equal energy $\sim E/2$.  This process gives a subleading contribution in comparison with that in Fig.~\ref{fig4}.
}
\label{fig3}
\end{figure}

To estimate  the lifetime of a pseudo-spin with  a size much smaller than  $R_1(T)$ (and thus the energy much larger than $E_*$), we have to find the most efficient  interaction process that allows to flip this pseudo-spin. 
One type of  such processes involves  energy conversion  from a pseudo-spin with an energy $E$ to  
two pseudo-spins of approximately equal energy:  $E = E_1+E_2$, where $E_1 \simeq E_2 \simeq E/2$, see Fig.~\ref{fig3}. 
pseudo-spins of energy $\simeq E/2$ get in turn a finite life time due to decay in pseudo-spins of energy $\simeq E/4$, and so on, until the energy $E_*$ is reached. In Appendix \ref{decay} we calculate the decay rate of a pseudo-spin with energy $E$ due to this type of processes. It turns out that this channel of decay is very slow and  plays no role for energy transport.

The most efficient  way of  energy transfer  involves two pseudo-spins with high energies $E,\ E-\omega \gg E_*$
that are separated by an energy difference corresponding to the optimal network,   $|\omega| \sim E_*$.
In this case the set of delocalized states of the optimal network acts as a bath that assists the energy transfer  between 
pseudo-spins in the network with energy $E_*$.  
The typical sizes of the high-energy pseudo-spins $R_1\sim (t/E)^{1/\alpha}$ are approximately equal; they are much shorter than the  size of the optimal pseudo-spins, $R_1 \ll R_1(T)$, see Fig.~\ref{fig4}.
The matrix element $A$ of the pseudo-spin flip-flop process is estimated as (see Appendix \ref{matrix-element})
\begin{equation}
A \sim \frac{1}{\omega} \frac{V^2}{r_{12}^\beta r_{13}^\beta},
\label{matrix-element-high-energy}
\end{equation}
where $\omega \sim E_*$ and the typical distances between the high-energy spins and between the high- and low-energy spins are 
\begin{equation}
\label{r12}
r_{12}\sim \left(\frac{V}{t}\right)^{1/\beta}\frac{R_1(T)}{R_1}
R_1^{\alpha/\beta}(T)
\end{equation}
and $r_{13} \sim R^*_2(R_1(T))$, respectively. When obtaining Eq.~(\ref{r12}), we used the expression (\ref{ps-density}) for the density $\rho_{\rm ps}(R_1)$ of $R_1$-pseudo-spins.
The density of those pseudo-spins whose energy is within the window of the width $E_*$ around $E$ is thus $(E_*/E)\rho_{\rm ps}(R_1)$, which yields Eq.~(\ref{r12}).

\vspace{0cm}

\begin{figure}[ht]
\vspace{0cm}
\includegraphics[width=0.4\textwidth]{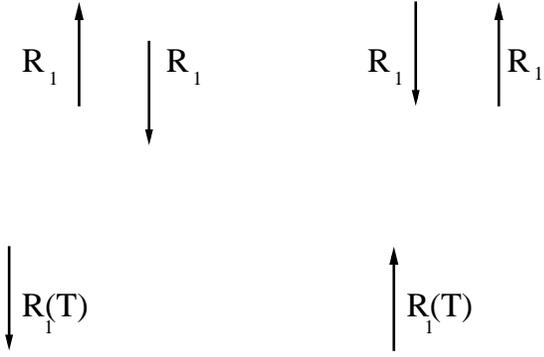} 
\vspace{0.2cm}
\caption{Dominant process responsible for delocalization of networks of high-energy pseudo-spins. This is a ``flip-flop'' process of two high-energy (small $R_1$) spins assisted by an excitation of the optimal network [$R_1(T)$]. 
}
\label{fig4}
\end{figure}
\vspace{0cm}

The typical transition  rate can be found  using the Fermi golden rule\cite{note-spin-bath}:
\begin{equation}
\label{tau}
\frac{1}{\tau} \sim \frac{A^2}{\Delta}\,,
\end{equation}
where $\Delta \sim E_*$ is  the characteristic level spacing of the bath (optimal network)  pseudo-spins  per volume with the linear size $\sim R_1(T)$.
Employing Eqs.~(\ref{matrix-element-high-energy}),  (\ref{r12}), and (\ref{tau}), we find
\begin{equation}
\frac{1}{\tau(R_1)}\sim E_*
\left(\frac{R_1}{R_1(T)} \right) ^{2\beta}.
\label{tau1}
\end{equation}
The contribution of these processes to  the thermal conductivity  is given by  
\begin{equation}
\label{kappa-integral1}
\kappa_{\rm loc} \sim \frac{1}{T^2}\int_{R_{1, {\rm th}}(T)}^{R_1(T)} \frac{dR_1}{R_1} \nu(R_1) E^3(R_1) \frac{r_{12}^2(R_1)}{\tau(R_1)}\,.
\end{equation}
Here $\nu(R_1)$ is  a density of states for pseudo-spins of the size $R_1$ with the energy $E(R_1)\sim t/R_1^\alpha$, and 
 $R_{1, {\rm th}}(T)$ is the thermal pseudo-spin size defined in Eq. (\ref{R-thermal}).  
 Using 
 $$
 \nu(R_1) \sim \frac{\rho_{\rm ps}(R_1)}{E(R_1)} \sim \frac{1}{r_{12}^d} \frac{E(R_1)}{E_*} \frac{1}{E(R_1)} = \frac{1}{r_{12}^d} \frac{1}{E_*},
 $$
 we obtain, after a straightforward algebra,
\begin{equation}
\label{kappa-integral2}
\kappa_{\rm loc} \sim \kappa_* \int_{R_{1, \rm th}(T)}^{R_1(T)}\frac{dR_1}{R_1}\left(\frac{R_1}{R_1(T)} \right)^{d-2-3\alpha+2\beta}.
\end{equation}
Here $\kappa_*$ is the thermal conductivity of the optimal network  (\ref{kappa-optimal})  that scales with temperature as $T^\mu_{*}$ with the exponent $\mu_{*}$ given by Eq.~(\ref{exp-mu}).

It is easy to see that
\be 
d-2-3\alpha+2\beta <0,
\label{condition-thermal}
\ee
in the entire parameter region $\alpha>\beta>d$. Thus,  the integral (\ref{kappa-integral2}) is dominated by its lower limit, which implies that
 the energy transport  by thermal pseudo-spins is more efficient than the one by the optimal network.  The resulting heat conductivity 
is given by
\begin{equation}
\kappa \sim \kappa_* \left  (\frac{R_{1, \rm th}(T)}{R_1(T)}\right)^{d-2-3\alpha+2\beta}
\label{kappa-thermal}
\end{equation}
and scales with temperature as 
\be
\label{kappa1}
\kappa \propto T^{\mu},
\ee
where
\begin{eqnarray}
\label{mu}
\mu & = & \mu_{*} - \delta, \nonumber\\
\delta & = & \frac{2\alpha\beta-d(\alpha+\beta)}{\alpha[(\alpha+\beta)d - \alpha\beta]}(3\alpha-2\beta-d+2).
\end{eqnarray}

In the  physically relevant situation of $d=2$ and $\alpha=\beta=3$ 
(which, in particular, corresponds to a 2D Anderson insulator with Coulomb interaction), Eqs. (\ref{kappa-thermal}) and (\ref{kappa-d2}) yield the exponent
\begin{equation}
 \mu = 5
 \end{equation}
 and the thermal conductivity
 \begin{equation}
 \kappa \sim \frac{t^4 V^4 \rho^{12}}{W^{12}}T^5.
 \label{kappa-thermal-d2}
\end{equation}

Let us now turn to the case $\alpha=\beta=d$, with $d=3$ having important physical applications. In this case, we are on the border of the regime set by inequalities $\alpha,\beta > d$, see Eq.~(\ref{conditions-alpha-beta}). As has been already mentioned in Sec.~\ref{section-Resonant-spin-networks}, the non-interacting system in such a situation is critical \cite{Anderson1958,Levitov90,Levitov,Mirlin_Evers} and exhibits an anomalous diffusion  of the form $r\sim t^{1/d}$. For $d=3$ (or, more generally, $d>2$) this is a subdiffusion, which is not sufficient to yield a nonzero DC transport coefficient (thermal conductivity). Therefore, the interaction-induced mechanism of establishing thermal transport, as explored in this work, retains its importance in such a situation as well. It follows from Eq.~(\ref{R1}) that for $\alpha=\beta=d=3$ the optimal pseudo-spin size $R_1(T)$ scales with temperature as $T^{-1/3}$, i.e., in the same way as $R_{1, {\rm th}}$. However, under the assumption of relatively small density of spins, $\rho^2Vt/W^2\ll 1$, 
these two scales remain different, 
\be
\frac{R_1(T)}{R_{1, \rm th}} \sim \left(\frac{\rho^2 Vt}{W^2}\right)^{-1/3} \gg 1. 
\label{scales-ratio-d3}
\ee
 Again,  the dominant contribution to transport is provided by thermal excitations. This yields the thermal conductivity (\ref{kappa-thermal}) with $\kappa_*$ given by Eq.~(\ref{kappa-d3}).  Using Eq.~(\ref{scales-ratio-d3}), we find
 \begin{equation}
\label{kappa-thermal-d3}
\kappa \sim V^{7/3} \left(\frac{t \rho^2}{W^2}\right)^{8/3}T^{4/3}\,,
\end{equation} 
which has the same temperature dependence as the optimal-network contribution (\ref{kappa-d3}) but an enhanced prefactor.

 \section{Thermal transport: Low-energy excitations}
\label{section-Thermal-transport-Low-energy-excitations}
 
Finally,  we estimate the transport via network of large pseudo-spins, with a typical size $R_1 \gg R_1(T)$.  
Each one of them is connected to a large number of partners, $N_2(R_1,R_2^*(R_1)) \gg 1$, which sets the size of the random-matrix blocks in Fig.~\ref{fig2}.
Therefore, the characteristic excitation energy  is $E(R_1)=N_2^{1/2}[R_1,R_2^*(R_1)]t/R_1^\alpha$.
The corresponding contribution to the heat conductance is given by Eq.~(\ref{kappa-r1}). 
Preforming straightforward algebraic calculations,  we find
\begin{eqnarray}
\kappa(R_1) &\simeq& \frac{t^{11/2}\rho^5T^{1/2}}{W^5}\left(\frac{V}{t}\right)^{(4+3d)/2\beta} \nonumber \\
& \times & R_1^{[\alpha(4+3d)+5d\beta-11\alpha\beta]/2\beta}.
\end{eqnarray}
The total contribution of low-energy dipoles (the ``infrared contribution'') to the thermal conductivity is obtained by summing over networks
with $R_1 \gg R_1(T)$,
\be
\label{kappa-low-energy}
\kappa_{\rm IR} \sim \int_{R_1(T)}^\infty \frac{dR_1}{R_1} \kappa(R_1).
\ee
This contribution is infrared-convergent if
\begin{equation}
d < \frac{\alpha(11\beta-4)}{3\alpha+5\beta},
\label{condition-low-en}
\end{equation}
or, equivalently, 
\begin{equation}
\label{condition-low-en2}
 \beta>\frac{\alpha(3d+4)}{11\alpha-5d},
\end{equation}
and divergent otherwise. For physically interesting situations the condition (\ref{condition-low-en}) is fulfilled, so that big dipoles do not play any important role. In particular,
for $\alpha=\beta =3$  the inequality  (\ref{condition-low-en}) amounts to $d<29/8$, which is fulfilled in view of Eq.(\ref{conditions-alpha-beta}). 

On the other hand, if we consider the whole range of parameters $\alpha$, $\beta$, $d$, satisfying the conditions (\ref{conditions-alpha-beta}), we find some regions where the inequality (\ref{condition-low-en}) is violated. More precisely, for $d<4/3$ this happens in a small  part of the II$_b$ region of the parameter space adjacent to the point $\alpha=\beta=d$. 
As an example, for $d=1$ and $\alpha=\beta$ this is the case  in the interval $1<\alpha<12/11$. In this situation, the thermal conductivity diverges due to very efficient propagation of low-energy modes. 
The situation is somewhat analogous to those encountered in phonon energy transport  
in dielectrics \cite{Levinson} and plasmon energy transport in disordered Luttinger liquid \cite{Fazio}. As has been already stated, we are not aware of any physical realization of such a regime in the present problem.

At this point, we remind the reader that there is an alternative potential mechanism for an infrared divergence of the thermal conductivity. These are the power-law tails that have been neglected in Eq.~(\ref{multistep-distribution}). We will analyze their effect in the next Section.

 \section{Power-law tails and Levy flights}
\label{section-tails}

In the previous Sections, we have analyzed the transport in a set of pseudospin netwrorks with the approximation Eq.~(\ref{multistep-distribution}) for the pseudospin interaction. As was mentioned below Eq.~(\ref{multistep-distribution}), the power-law tails discarded there
[interactions at distances larger than $R_2^*(R_1)$]
 may potentially lead to a divergent contribution to the thermal conductivity. In this Section, we will analyze under what condition this mechanism of superdiffusive energy transport becomes operative. 

Let us consider the optimal network (or any other conducting network) and include the power-law tails perturbatively. Since the states on the network are delocalized (and thus broadened), we can consider the effect of the corresponding long-distance hopping processes on a classical level (i.e. without looking for resonances). The probability of a jump to a distance $r$ will then be proportional to the squared absolute value of the corresponding matrix element $V/r^\beta$.  
 The contribution of these processes to the thermal diffusion coefficient (and thus to thermal conductivity) will thus be given by the following integral over $r$:
\begin{equation}
\kappa_{\rm tail} \propto \int d^d r\:   r^2 |V/r^\beta|^2 \propto \int d^d r\:  r^{2-2\beta}.
\end{equation}
Here we have only kept powers of $r$, since we are only interested in the possible infrared divergence of the $r$ integration. 
The condition for infrared  convergence of this integral is $d < 2\beta  - 2$, or, equivalently, 
\begin{equation}
 \beta > 1 + d/2 .
 \label{kappa-tail}
\end{equation}
If this condition is not fulfilled, the thermal conductivity is infinite. 

It is easy to check that that in the whole region of our interest, $\alpha \ge d$, the condition (\ref{condition-low-en2}) follows from the inequality (\ref{kappa-tail}). Therefore, the power-law-tail mechanism of the infrared divergence of the thermal conductivity (studied in this Section) is more efficient than the one due to ultra-low-energy networks (considered in Sec.~\ref{section-Thermal-transport-Low-energy-excitations}). 

The inequality (\ref{kappa-tail}) is always satisfied in our problem in spatial dimensionalities $d \ge 2$, since we assume $\beta \ge d$ (and the strict inequality $\beta > 2$ for $d = 2$). If, however, the spatial dimensionality $d$ is lower than two, a region emerges,  
 $\beta < 1 + d/2$,  where the thermal transport is of superdiffusive (Levy-flight) character. Specifically, the displacement then scales with time as $r \sim t^{1/z}$, with the dynamical exponent $z=2\beta-d$.   In particular, for $d = 1$ the superdiffusion occurs under the condition $\beta < 3/2$.

 \section{Pseudo$^2$-spin networks and thermal transport}
 \label{section-Pseudo2-spin-networks-and-thermal-transport}
 
 In Secs.~\ref{section-Thermal-transport-Optimal-network}, \ref{section-Thermal-transport-High-energy-excitations}, and \ref{section-Thermal-transport-Low-energy-excitations} we have presented a detailed analysis of the thermal conductivity due to pseudo-spin networks.
In the present Section we consider the energy transport at the next level of hierarchical construction, i.e., the transport via 
 pseudo$^2$-spin networks. As has been explained in Sec.~\ref{section-Resonant-spin-networks3}, this mechanism is operative in the regions III and II$_{\rm c}$ of the phase diagram of Fig.~\ref{Fig:PhaseDiagram}. In the region III the pseudo$^2$-spins provide the only mechanism of transport. The situation is more intricate in region II$_{\rm c}$ where the connected pseudo-spin and pseudo$^2$-spin networks coexist, as has been already discussed in the end of Sec.~\ref{section-Resonant-spin-networks3}. We will return to the question of implications of the existence of delocalized pseudo$^2$-spin networks for the transport phenomena in region II$_{\rm c}$ in the end of this Section. 
 
 The logics of the analysis of the thermal transport due to pseudo$^2$-spins is essentially the same as for pseudo-spins (Secs.~\ref{section-Thermal-transport-Optimal-network}, \ref{section-Thermal-transport-High-energy-excitations}, and \ref{section-Thermal-transport-Low-energy-excitations}): we will first calculate the contribution to the thermal conductivity due to the optimal connected network and then will analyze contributions of high-energy and  low-energy pseudo$^2$-spins. In view of this similarity, we keep the exposition in this Section relatively concise.
  
 In analogy with Sec. \ref{section-Thermal-transport-Optimal-network}, we first consider the thermal conductance due to the optimal pseudo$^2$-spin network defined by the condition
 \begin{equation}
  N_3(R_{1, \rm th}, R_2, R_2)\sim 1,
 \end{equation}
which determines, together with Eq. (\ref{N3*}),  the optimal size of pseudo$^2$-spins,
\begin{eqnarray}
 R_2(T)&\sim&\left(\frac{W^4}{tV\rho^4 T^2}\right)^{1/(2d-\beta)}R_{1, {\rm th}}^{(\alpha-2d)/(2d-\beta)}\nonumber\\
 &\sim&\left(\frac{W^4}{V\rho^4}\right)^{1/(2d-\beta)} t^{-\frac{2d}{\alpha(2d-\beta)}}T^{-\frac{3\alpha-2d}{\alpha(2d-\beta)}}. \hspace{0.5cm}
\label{R2-optimal}
 \end{eqnarray}
It is easy to see that at low temperatures  $R_2(T)\gg R_{1, \rm th}(T)$ as long as $\alpha >2d$ and $\beta < 2d$, which is the case in the regions II$_{\rm c}$ and III of the phase diagram. 

To determine the temperature scaling of the thermal conductivity due to optimal pseudo$^2$-spin network, we use the analogue of Eq. (\ref{kappa-optimal}). Substituting there the  characteristic energy of excitations on the optimal pseudo$^2$-spin network, $E^{(2)}_{*}\sim V/R_2^\beta(T)$, we find
\begin{equation}
 \kappa^{(2)}_*\sim \frac{\left[E_*^{(2)}\right]^3}{T^2} \left[R_2(T)\right]^{2-d}
 \propto T^{\nu_*},
 \label{kappa2_opt}
\end{equation}
with the exponent
\begin{equation}
 \nu_*=\frac{(3\alpha-2d)(3\beta+d-2)}{\alpha(2d-\beta)}-2.
 \label{kappa2*}
\end{equation}

We now include into consideration the high-energy pseudo$^2$-spins and study the energy exchange between two such pseudo$^2$-spins assisted by the delocalized modes in the optimal network (cf. Sec \ref{section-Thermal-transport-High-energy-excitations}). 
The typical distance over which the energy is transfered in such a process is 
\begin{equation}
 r_{12}=\left(\frac{1}{\rho_{{\rm p}^2{\rm s}}(R_1, R_2)R_2^\beta/R_2^\beta(T)}\right)^{1/d},
\end{equation}
 where $\rho_{{\rm p}^2{\rm s}}(R_1, R_2)$ is the density of pseudo$^2$-spins given by Eq. (\ref{p2s-density}). Under the assumption $R_1<R_1^*(R_2)\equiv (t R_2^\beta/V)^{1/\alpha}$, we have
 \begin{eqnarray}
&&  \rho_{{\rm p}^2{\rm s}}(R_1, R_2) \sim \frac{\rho^4 tV T^2}{W^4}R_1^{2d-\alpha}R_2^{d-\beta}\nonumber\\
 && \hspace{1cm} \sim \frac{1}{R_2^d}\left(\frac{R_{1, \rm th}(T)}{R_1}\right)^{\alpha-2d}\left(\frac{R_2}{R_2(T)}\right)^{2d-\beta}. \hspace{1cm}
 \end{eqnarray}
Thus,
\begin{equation}
r_{12}\sim
\frac{R^2_2(T)}{R_2}\left(\frac{R_1}{R_{1, \rm th}(T)}\right)^{(\alpha-2d)/d}.
\label{r12-2}
\end{equation}
Further, the matrix element for the pseudo$^2$-spin flip-flop process is estimated as [cf. Eq. (\ref{matrix-element-high-energy})]
\begin{equation}
A\sim \frac{1}{\omega}\frac{V^2}{r_{12}^\beta r_{13}^{\beta}}\sim \frac{V}{r_{12}^{\beta}},
\label{A2}
\end{equation}
and  the density of states in the optimal network in a volume with a linear size $R_2(T)$ is given by 
\be
\label{Delta2}
1/\Delta\sim 1/E_*^{(2)}\sim R_2^\beta(T)/V.
\ee
Substituting Eqs.~(\ref{r12-2}), (\ref{A2}), and (\ref{Delta2}) in the golden-rule formula (\ref{tau}), we find the characteristic lifetime of the high-energy pseudo$^2$-spins, 
\begin{equation}
\frac{1}{\tau}\sim \frac{V}{R_2^\beta(T)}\left(\frac{R_2}{R_2(T)}\right)^{2\beta}\left(\frac{R_{1, \rm th}(T)}{R_1}\right)^{2\beta (\alpha-2d)/d}.
\label{tau2}
\end{equation} 
The contribution of the high-energy pseudo$^2$-spin excitations to the heat conductivity is thus
\begin{widetext}
\begin{eqnarray}
\kappa^{(2)}_{\rm loc}&\sim &\frac{1}{T^2}\int_{R_{2, \rm th}(T)}^{R_2(T)}\frac{dR_2}{R_2}
\int_{R_{1, \rm th}(T)}^{R_1^*(R_2)}\frac{dR_1}{R_1}\nu(R_1, R_2)E^3(R_2)\frac{r_{12}^2(R_1, R_2)}{\tau(R_1, R_2)}\nonumber\\
&\sim & \frac{V^3}{T^2}\left[R_2(T)\right]^{2-d-3\beta}\int_{R_{2, \rm th}(T)}^{R_2(T)}\frac{dR_2}{R_2}
\int_{R_{1, \rm th}(T)}^{R_1^*(R_2)} \left(\frac{R_{1, \rm th}(T)}{R_1}\right)^{(\alpha-2d)(d+2\beta-2)/d}
\left(\frac{R_2(T)}{R_2}\right)^{\beta-d+2}.
\label{kappa_loc2}
\end{eqnarray}
\end{widetext}
Here we have taken into account that the characteristic energy of pseudo$^2$-spins of size $R_2$ is
$E(R_2)\sim V/R_2^\beta$ and introduced the short-distance cutoff for the $R_2$ integration, $R_{2, \rm th}(T)\sim (V/T)^{1/\beta}$.

It is easy to see that for $d>2/3$ and for $\alpha$ and $\beta$ within regions III and II$_{\rm c}$ of the phase diagram, the integral in Eq. (\ref{kappa_loc2}) is dominated by short distances. Correspondingly, 
the energy transport by high-energy pseudo$^2$-spins dominates over that by the optimal pseudo$^2$-spin network, and the thermal conductivity of the system is given by
\begin{equation}
\kappa^{(2)}\sim \kappa^{(2)}_{*} \left(\frac{R_2(T)}{R_{2, \rm th}(T)}\right)^{\beta-d+2}\sim T^\nu,
\label{kappa2}
\end{equation}
with the exponent 
\begin{equation}
\nu= \nu_*-2\frac{[2\alpha\beta-d(\alpha+\beta)](\beta-d+2)}{\alpha\beta(2d-\beta)}.
\label{nu2}
\end{equation}

Let us finally discuss the third possible transport mechanism in the system: transport via pseudo$^2$-spins of very low energies [i.e., with sizes $R_2\gg R_2(T)$] that build networks with   $N_3(R_{1, \rm th}, R_2, R_2)\gg 1$.  A consideration analogous to the one in Sec. \ref{section-Thermal-transport-Low-energy-excitations} leads to the estimate
\begin{equation}
\kappa_{IR}\sim  \int_{R_2(T)}^{\infty}\frac{dR_2}{R_2}\kappa(R_2)
\end{equation}
with
\begin{eqnarray}
\kappa(R_2) &\sim& \frac{V^3}{T^2}\frac{1}{R_2^{3\beta+d-2}}N_3^{5/2}(R_{1, \rm th}, R_2, R_2) \nonumber \\
&\sim& R_2^{(8d+2-11\beta)/2}.
\end{eqnarray}
Since $8d+2-11\beta<0$ for all $\beta>d> 4/7$, the ultra-low energy networks of pseudo$^2$-spins are not important for the transport in physically relevant spatial dimensions ($d\ge 1$). 

The analysis of power-law tails ($\sim 1/r^\beta$) of the interaction performed in Sec.~\ref{section-tails} for the case of a pseudo-spin network fully applies to delocalized pseudo$^2$-spin networks. Specifically, under the condition (\ref{kappa-tail}) no infrared divergence of the thermal diffusion 
constant occurs; otherwise, the transport is superdiffusive.  

The thermal conductivity due to the pseudo$^2$-spin network, Eqs.~(\ref{kappa2}) and (\ref{nu2}), constitutes the main result of this Section.
This mechanism controls the thermal transport in the region III of the phase diagram, so that Eqs.~(\ref{kappa2}) and (\ref{nu2}) yield the final result for the thermal conductivity in this region.  The pseudo$^2$-spin mechanism of thermal transport is also relevant in region II$_{\rm c}$ but there the physics is more involved, in view of coexistence of pseudo-spin and pseudo$^2$-spin networks, see Sec.~\ref{section-Resonant-spin-networks3}. The first idea then would be simply to consider them as parallel transport channels and to add the corresponding contributions to the thermal conductivity, i.e.the pseudo-spin contribution (\ref{kappa1}),  (\ref{mu}) and  the pseudo$^2$-spin thermal conductivity (\ref{kappa2}), (\ref{nu2}). The larger of the two contributions would then win.  We expect, however, that in a part of the region II$_{\rm c}$ the situation may be still more intricate. Specifically, 
the pseudo-spin transport mechanism described in  Sec. \ref{section-Thermal-transport-High-energy-excitations} relies on flip-flop processes of thermal pseudo-spins assisted by the optimal pseudo-spin network. The corresponding time scale is given by 
Eq. (\ref{tau1}) and diverges at the line $d(\alpha+\beta)=\alpha\beta$, which is the  border of the II$_{\rm c}$ regime. 
The connected pseudo$^2$-spin network will provide an alternative decay channel for the thermal pseudo-spins. The corresponding time scale, $\tau^\prime$, is expected to be longer  than the  time (\ref{tau2}) (which controls  the decay of  {\it pseudo$^2$-spins }). However,   the time $\tau^\prime$ will stay finite at the line $d(\alpha+\beta)=\alpha\beta$. Thus, in a part of the region II$_{\rm c}$ close to this line, the pseudo-spin relaxation time will be determined by pseudo$^2$-spins. Therefore, the contribution of pseudo-spins to the thermal conductivity will be determined by a mixed transport mechanism---flip-flops of thermal pseudo-spins assisted by optimal pseudo$^2$-spin network. We expect that this transport mechanism may give a dominant contribution to thermal conductivity (and will thus control its temperature scaling) in a certain part of the region II$_{\rm c}$ near the boundary with the region III.   We do not explore this question in the present paper, leaving it as an interesting direction 
for future research.

We remind the reader that, as was pointed out in Sec. \ref{s21}, the results for the thermal conductivity in Sec. \ref{section-Thermal-transport-Optimal-network}---\ref{section-Thermal-transport-Low-energy-excitations}
and \ref{section-Pseudo2-spin-networks-and-thermal-transport} were obtained within the approximation that neglects the spectral diffusion. In the remaining part of
the paper, we take the spectral diffusion in consideration and analyze its implications.

  \section{Localization threshold}
  \label{section-Localization-threshold}
 
While the main goal of this paper is the analysis of the thermal transport,
in the present Section we discuss another related  aspect of the problem. Specifically, we will analyze the system-size scaling of the total energy (with the ground-state energy set to zero) representing the localization threshold for many-body states \cite{Burin15}. 
To set the stage for this discussion, let us first remind the reader about two known types of such scaling in fermionic many-body systems:

\begin{itemize}

\item[(i)]
 If there is a delocalization transition in the non-interacting system, the corresponding threshold $E_c$ will be independent of system size $L$ in the limit $L\to\infty$. This is the most conventional case of mobility edge of the Anderson transition. 
 
 \item[(ii)]
 If all single-particle states are localized and the interaction is of short-range character, the system may undergo a transition between the low-temperature localized phase \cite{fleishman80} and the high-temperature delocalized phase \cite{altshuler81} at a certain critical temperature $T_c$, see Refs.~\onlinecite{gornyi05,basko06,luitz15,nandkishore15} for analytical predictions and numerical simulations, as well as Refs.~\onlinecite{schreiber15, bloch16} for experimental realization of the transition in 1D and 2D cold-atom systems. In this situation, a many-body state is delocalized if its energy is above the threshold $E_c \propto T_c^2 L^d$, which scales as $L^d$ with the system size.

 \item[(iii)] An interacting electronic system  in a quantum dot is described by a Hamiltonian characterized by the single-particle mean level spacing $\Delta \propto L^{-d}$ and the typical value $\Delta/g$ of the matrix element of interaction, where $g$ is the dimensionless conductance. 
 For given value of $g\gg 1$ (viewed as an independent parameter), the system undergoes the localization transition in the Fock space at the threshold  energy $E_c\propto L^{-d}$, see Refs. \onlinecite{AGKL,jacquod97,mirlin97,silvestrov98,GornyMirlinPolyakov2016}.

 \end{itemize}

 Let us show that the system studied in the present work exhibits a behavior which is intermediate  with respect to the  cases (ii) and (iii) above.
 Indeed, let us consider our system at a certain temperature $T$ in a box of finite size $L$. We begin by considering the situation in which the delocalization is governed by  pseudo-spin resonances. (The corresponding result will be applicable in the region II$_{\rm b}$ and potentially in a part of the region II$_{\rm c}$.)
 We know that the delocalization is achieved due to coupling of pseudo-spins at a distance $R_2^*(R_1(T))$. If the system size $L$ is reduced to a value much smaller than this distance, no resonance couplings between pseudo-spins, and thus no delocalization will take place.
 We thus find the following condition for the localization threshold:
\begin{equation}
\label{threshold}
R_2^*(R_1(T_c)) \sim L\,,
\end{equation}
where $R_2^*$ is given by Eqs.~(\ref{R2star}) and (\ref{R1}), which yields the scaling of the critical temperature $T_c$ with the system size  and disorder
\begin{equation}
T_c\propto W^{2}L^{-\frac{(\alpha+\beta)d-\alpha\beta}{\alpha}}\,.
\label{Tc}
\end{equation}
This result  can be  translated into  critical energy
\begin{equation}
E_c \propto L^d \frac{T_c^2}{W} \propto W^3 L^{\frac{2\beta(\alpha -d)}{\alpha}-d}\,,
\label{Ec}
\end{equation}
which yields the minimal energy of a delocalized excitation in a finite-size system at zero temperature.  It is easy to check that, by virtue of inequalities (\ref{conditions-alpha-beta}) and (\ref{condition-deloc}), the exponent $\gamma$ determining the scaling of $E_c$ with $L$,
\be
\gamma = \frac{2\beta(\alpha -d)}{\alpha}-d,
\label{gamma}
\ee
 satisfies $-d<\gamma<d$
everywhere in the regions II$_{\rm b}$ and II$_{\rm c}$,  so that we are indeed in a situation intermediate between the above cases (ii) and (iii).

 An important question which we are now going to address is the character of the many-body state on the delocalized side of the transition. To be specific, let us fix temperature $T$ and study the delocalization (e.g.,  the evolution of level statistics for many-body states) upon increase of the  system size $L$. 
For $L$ smaller than the delocalization length $L_c(T)\equiv R_2^*(R_1(T))$ the statistics is clearly of Poisson form since all degrees of freedom are localized.  As discussed above,  at $L\approx L_c$ the first resonant pair of spins of optimal size $R_1(T)$ appears. How does the delocalization  proliferate upon further increase of $L$?  We argue now that already at the length of a few $L_c$ (where  ``a few'' means above some critical number of order unity) all spin degrees of freedom become delocalized. Indeed, in such a system we have a few coupled resonant spin pairs (i.e., a few pseudospins). Flipping any of them yields a new many-body state that is well connected (matrix element larger or of the order of the energy splitting) with the original one. This state again possesses a few coupled resonant spin pairs and thus is well connected with a few other many-body states, etc. We argue now that this yields a Bethe-lattice structure. The key point is that resonances are efficiently  ``eliminated''  by flipping other pairs. Specifically, consider a certain resonant spin pair 1 that existed in the original state. If we flip another spin pair 2, the energy of pair 1 is shifted by $\sim E_*$. After $p$ steps, we will flip $p$ pairs, and the spins participating there will be distributed in space roughly uniformly within the  length $L_c$. The shift of the energy of the pair 1 will be determined, in view of $\beta > d$, by the $S^z S^z$ interaction  [see discussion after Eq. (\ref{res})] with the spin closest to one of spins within the pair 1. The corresponding distance is $\sim L_c p^{-1/d}$. This yields the shift 
\begin{equation}
\Delta^{(p)} E \sim E_* p^{\beta/d},
\label{e1}
\end{equation}
an thus the return probability $\sim p^{-\beta/d}$ to the resonance window of width $E_*$.
Since $\beta > d$, the ``return probability'' obtained by summing $\sim p^{-\beta/d}$ over $p$, converges \cite{note-return-prob}. Thus, with the probability of order unity, the pair 1 never returns to the set of resonance pairs. Therefore, it appears to be a good approximation to consider the emergent structure in the many-body Hilbert space as a Bethe lattice. Clearly, the Bethe-lattice approximation works only until all available (thermal) many-body spin states are exhausted; beyond this generation the effective lattice in the many-body space gets ``compactified''. In this sense, one may approximately view the effective lattice in the Fock space of the system as a tree-like structure without the boundary, such as a random regular graph \cite{tikhonov16}.

This argument implies that the length $L_c$ marks the many-body localization transition for our finite system  (``spin quantum dot''). In particular, the statistics of many-body excitations is Poisson on one side and Wigner-Dyson on the other side of the transition. Of course, the transition becomes sharp only in the limit of large number of involved spins, 
\begin{equation}
N_s \equiv L_c^d \rho T/W \gg 1,
\label{e2}
\end{equation}
which is fulfilled in view of our assumptions of low temperature and strong disorder, $T \ll W$ and $t\rho^{\alpha/d} \ll W$. 

Some available numerical studies support these expectations. In particular,  in Ref. \onlinecite{Burin15b}  a numerical analysis of the finite-size delocalization transition was performed for the case of infinite $T$ for $d=1$ and $\alpha=\beta$. The infinite-$T$ limit corresponds to setting $T \sim W$ in Eq.~(\ref{Tc}), which yields, for $d=1$ and $\alpha=\beta$, the critical disorder $W_c \propto L^{2-\alpha}$. Numerical results of Ref. \onlinecite{Burin15b} supported the existence of a delocalization transition (which became sharper with increasing $L$) around this $W_c$ .

We turn now to the similar analysis for the case when the delocalization is due to pseudo$^2$-spin resonances. The condition for the localization threshold then reads, in analogy with Eq.~(\ref{threshold}), 
\begin{equation}
\label{threshold-2}
R_2(T_c) \sim L\,,
\end{equation}
where $R_2(T)$ is the size of the optimal pseudo$^2$-spin given by  Eq.~(\ref{R2-optimal}). This yields 
the scaling of the critical temperature $T_c$ with the system size and disorder strength \cite{FootNoteLogBette},
\begin{equation}
T_c\propto W^{\frac{4\alpha}{3\alpha-2d}}L^{-\frac{\alpha(2d-\beta)}{3\alpha-2d}}\,,
\label{Tc2}
\end{equation}
or, equivalently, the  critical energy
\begin{equation}
E_c \propto \frac{L^d T_c^2}{W} \propto W^{\frac{5\alpha+2d}{3\alpha-2d}}L^{\gamma_2}\,,
\label{Ec2}
\end{equation}
with the exponent $\gamma_2$ governing the temperature scaling,
\be
\gamma_2 = \frac{\alpha\beta + \alpha d - 2d^2}{3\alpha-2d}.
\label{gamma2}
\ee
This result is applicable in the region III. Furthermore, the pseudo$^2$-spin delocalization mechanism is operative also in the region II$_{\rm c}$ where it competes with the pseudo-spin delocalization. In order to find out, which of the two mechanisms determines the delocalization threshold in this region, we compare the exponents $\gamma$ and $\gamma_2$. It turns out that $\gamma_2 < \gamma$ in the whole region  II$_{\rm c}$.  Therefore, in addition to region III, the pseudo$^2$-spin mechanism determines the delocalization threshold also in the region II$_{\rm c}$, with the critical energy given by Eq.~(\ref{Ec2}).  The exponent $\gamma_2$ satisfies $-d<\gamma_2<d$ everywhere in the regions III and II$_{\rm c}$, again implying a situation intermediate between the cases (ii) and (iii) described in the beginning of this Section. The exponents $\gamma$ and $\gamma_2$ match at the border $\alpha=2d$ of the regions  II$_{\rm b}$ and II$_{\rm c}$. Further, the exponent $\gamma_2$ takes the value $d$ at the boundary 
line $\beta=2d$, thus ensuring a matching with the scaling of the type (ii) characteristic for a many-body-localization threshold in Anderson insulators with short-range interaction.  

 It is worth mentioning that the delocalization of pseudo-spins at the corresponding threshold $E_c$ in the regions III and IIc does not yet imply the delocalization of spins: the latter will take place at a higher energy. This is a manifestation of the fact that the delocalization of different excitations in a finite-size system may take place at parametrically different energies.

\section{Spin relaxation and spectral diffusion}
 \label{section-Spin-relaxation-and-spectral-diffusion}
 
 In Sec. \ref{section-Localization-threshold} we have analyzed the many-body localization in a ``spin quantum dot''.   We are now going to study   the implications of the physics discussed there for the main subject of the present work---thermal transport in an extended system. 
 
 In  Secs. \ref{section-Thermal-transport-Optimal-network} --- \ref{section-tails} we assumed that the pseudo-spins (or pseudo$^2$-spins in the regions of II$_c$ and III of the phase diagram) can be viewed as rigid objects built out of specific spins that happened to be in resonance. All other spins were ignored. We drop this assumption from now on  and study the contribution of all spins to the thermal transport.   
 
 We focus first on the regions II$_b$ and II$_c$. 
 Let us consider a typical thermal spin. First, in complete analogy to Sec. \ref{section-Thermal-transport-High-energy-excitations},  its interaction with the optimal network leads to  spin relaxation via the process of simultaneous flip of two thermal spins  assisted by pseudo-spin flipping in the network. 
We do not present here the corresponding spin relaxation rate and the contribution of such processes to thermal transport (referring the interested reader to Supplemental Material \cite{Supplemental}) as it turns out that  there exist a faster channel for the spin relaxation and thermal transport. 
Specifically, as we have seen in Sec. \ref{section-Localization-threshold}, transitions of the resonant pseudo-spins shift energies of other spins, thus  destroying neighboring pseudo-spins and creating new ones. In a piece of the system of the size $L_c(T)=R_2^*(R_1(T))$, there is of order of one pseudo-spin whose flip occurs typically within the time $\sim 1/E_*$.
Thus the rate for relaxation of all spins (or, equivalently, for any given spin) can be  estimated as (cf. Ref.~\onlinecite{Burin94}).
\begin{equation}
1/\tau_{\rm sd} \sim E_* N_s^{-1}\sim t \frac{t\rho}{W} \left[R_1(T)\right]^{d-2\alpha}.
\label{tau_sd}
\end{equation}
Such a relaxation mechanism is known as spectral diffusion (thus the subscript ``sd'') in the   theory of spectral lines as measured in spin resonance experiments \cite{Klauder62} and was used to estimate the relaxation rate of two-level tunneling systems in amorphous solids in Refs. \onlinecite{Galperin83,Burin94}. 

The spin relaxation rate  (\ref{tau_sd}) is an important characteristics of the system.  For the dipole-dipole interaction, $\alpha=\beta=3$,  we find
\begin{eqnarray}
1/\tau_{\rm sd} \propto T W^{-3}\,, \qquad d=3,\label{tau_sd3} \\
1/\tau_{\rm sd} \propto T^4 W^{-9}\,, \qquad d=2.\label{tau_sd2}
\end{eqnarray}

Equation (\ref{tau_sd}) allows us to estimate the contribution of spectral diffusion to thermal transport. Indeed, coming into a resonance with another spin, a thermal spin transports  an energy of order $T$ over distance $R_1(T)$. Correspondingly [cf. Eq. (\ref{kappa-integral1})] the thermal conductivity is given by
\begin{multline}
\kappa_{\rm sd}=T\times \frac{\rho}{W} \times t \frac{t\rho}{W} \left[R_1(T)\right]^{d-2\alpha}\times \left[R_1(T)\right]^2\\=T\left(\frac{t\rho}{W}\right)^2 \left[R_1(T)\right]^{d-2\alpha+2}
\label{Eq:kappaSD}
\end{multline}
and scales with the temperature as
\begin{equation}
\kappa_{\rm sd}\propto T^{\mu_{\rm sd}}\,, \qquad \mu_{\rm sd}=1+\frac{\beta(d-2\alpha+2)}{(\alpha+\beta)d-\alpha\beta}.
\label{kappa_sd_T}
\end{equation}

It is instructive to compare  Eq. (\ref{Eq:kappaSD})  to the thermal conductivity of the optimal network, Eq. (\ref{kappa-optimal}), which yields 
\begin{equation}
\kappa_{\rm sd}=\left(\frac{T}{E_*}\right)^2 \times \left(\frac{R_1(T)}{R_2^*(R_1(T))}\right)^2 \times \kappa_*.
\label{relation-opt-sd}
\end{equation}
We see thus that the expression for the conductivity due to spectral diffusion can be obtained from the one due to optimal network by replacing the energy transfer $E_*$ by $T$ and the jump radius $R^*_2(R_1(T))$ by $R_1(T)$. In the particular case of $\alpha=\beta$ and $t=V$ we have $R_1(T) \sim R^*_2(R_1(T))$, so that only the modification of the energy transfer is needed.  The physical explanation of the relation (\ref{relation-opt-sd}) is as follows. In the picture of transport over the optimal network (which yields $\kappa_*$) we assumed that pseudospins are ``stable'' objects, and the allowed energy transfer processes are those between two pseudospins --  yielding the energy transfer $E_*$ and distance $R^*_2(R_1(T))$. Within the spectral diffusion argument, spins are constantly changing their resonant partners, so that the relevant energy transfer processes take place between spins, with the energy transfer $T$ and the distance $R_1(T)$. 
The spatial density of objects participating in the transport (pseudospins in the first picture or spins forming pseudospins in a given configuration in the second picture) is parametrically the same.  This immediately yields the relation (\ref{relation-opt-sd}). 

It can be checked (see Supplemental Material \cite{Supplemental})  that the conductivity (\ref{Eq:kappaSD}) dominates over the conductivity due to high energy pseudospins studied in Sec. \ref{section-Thermal-transport-High-energy-excitations} in the whole parameter ranges II$_{b}$ and II$_c$.  Thus, spectral diffusion provides the dominant  channel for the heat transport.

For $\alpha=\beta$ and $V \sim t$ Eq. (\ref{Eq:kappaSD}) takes a particularly simple form:
\begin{equation}
\kappa_{\rm sd}\sim t \left[R_1(T)\right]^{2-d-\alpha},
\end{equation}
and the temperature  scaling of thermal conductivity is  
\begin{equation}
\kappa_{\rm sd}\sim T^{\frac{\alpha+d-2}{2d-\alpha}}.
\end{equation}
In the physically most interesting case $\alpha=\beta=3$ we find, taking into account Eq. (\ref{R1}),
\begin{equation}
\kappa_{\rm sd}  \propto T^{4/3}W^{-8/3}
\label{kappa_sd3}
\end{equation}
for $d=3$ (cf. Ref.~\onlinecite{Burin89}) and
 \begin{equation}
\kappa_{\rm sd}  \propto T^{3}W^{-6}
\label{kappa_sd2}
\end{equation}
 for $d=2$.
 
 Comparing Eqs. (\ref{kappa_sd3}) and (\ref{kappa_sd2}) to Eqs. (\ref{kappa-thermal-d3}) and (\ref{kappa-thermal-d2}), we see that including the spectral diffusion into consideration is especially important in $d=2$ where it changes the temperature scaling of thermal conductivity from $T^5$ to $T^3$.  The effect is weaker in $d=3$ where the temperature scaling of conductivity remains unchanged and only the prefactor in enhanced by a factor of $(W/t)^{8/3}$. 
 
 Let us now  briefly analyze the effect of spectral diffusion on thermal transport in the parameter range III  where the pseudo$^2$-spin network  is responsible for delocalization. Taking into account that the spectral diffusion works now on the level of pseudo-spins and that in the optimal network $R_2\sim R_3$ we conclude that in this parameter range [cf. discussion after Eq. (\ref{relation-opt-sd})]
  \begin{equation}
 \kappa_{\rm sd}=\frac{T^2}{\left[E_*^{(2)}\right]^2}\kappa^{(2)}_*.
 \end{equation}
 Here $\kappa_*^{(2)}$ is the thermal conductivity of the pseudo$^2$-spin network, Eq. (\ref{kappa2_opt}) and $E^{(2)}_*=V/R^\beta_2(T)$ is the typical energy of its excitations. 
 Using Eq. (\ref{R2-optimal}), we find the temperature scaling of the thermal conductivity in the part III of the phase diagram
 \begin{equation}
 \kappa_{\rm sd}=V \left[R_2(T)\right]^{2-d-\beta} \propto T^{\frac{(3\alpha-2d)(\beta+d-2)}{\alpha(2d-\beta)}}.
 \label{kappa_2_sd}
 \end{equation}
 
 Before closing this Section, let us stress that the spectral diffusion picture (and thus the results of this Section) rely on the assumptions that correlations in shifts of energies of spins due to successive flips of resonant pseudospins as well as correlations between contributions of different ``paths'' in the many-body space can be neglected. While we have provided arguments in  favor of these approximations in Sec.~\ref{section-Localization-threshold} and in the present Section (see also Ref.~\onlinecite{Burin94}),  a more rigorous justification would be certainly of interest.

 \section{Summary}
 \label{section-summary}
 
In conclusion, we have studied  propagation  of energy  through  
the Anderson insulator  with  a long-range interaction. The system was described by the Hamiltonian (\ref{Hamiltonian}), with spins representing
particle-hole excitations formed by localized electronic states. While the $1/r$ Coulomb interaction between localized states leads to the dipole-dipole interaction between spins, with $\alpha=\beta=3$, we have considered $\alpha$ and $\beta$ as free parameters for generality, with the assumption $\alpha \ge \beta \ge d$. Resonant pairs of these spins were treated as pseudo-spin operators.  

The dominant channel of the heat propagation (and thus the scaling of the thermal conductivity) depends on relations between $d$, $\alpha$, and $\beta$.
Under the condition  (\ref{condition-deloc}), which defines regions II$_{\rm b}$ 
and II$_{\rm c}$ of the phase diagram in Fig.~\ref{Fig:PhaseDiagram},  the interaction between pseudo-spins leads to energy delocalization.   Specifically, excitations with energies below $E_*$, Eq.~(\ref{E-star}) [which corresponds to the pseudo-spin size (\ref{R1})] become delocalized by resonant couplings between pseudo-spins. As a consequence, pseudo-spins with higher energies can also exchange energy due to coupling to excitations with energy $\sim E_*$, which thus serve as a bath. The region II$_{\rm b}$ is of particular interest, as it contains the physically most relevant line $\alpha=\beta$.

If the condition (\ref{kappa-tail}) is violated and under, the thermal conductivity shows an infrared divergence, and the energy transport is of superdiffusive (Levy-flight) character. In the 1D case, this happens under the condition $\beta < 3/2$. 
This situation, however, is not realized in physically most interesting cases ($d\ge 2$).

If the condition (\ref{kappa-tail}) is fulfilled the dominant contribution to the thermal conductivity is provided by thermal excitation that move due to assistance of optimal ones.  
Under the approximation that neglects spectral diffusion, the  thermal conductivity is then given by Eq.~(\ref{kappa-thermal}), with the 
temperature dependence governed by the exponent   (\ref{mu}). Including the spectral diffusion in consideration increases the conductivity and leads to Eqs.~(\ref{Eq:kappaSD}) and~(\ref{kappa_sd_T}).

From the physical point of view, the cases of 2D and 3D systems with dipole interactions between spins, $\alpha=\beta=3$,  are of particular importance. These cases correspond to 2D and 3D Anderson insulators with the conventional ($1/r$) Coulomb interaction. 
For $d=2$ and $\alpha=\beta=3$, 
the obtained thermal conductivity scales with temperature as  $\kappa \propto T^3 $. 
This result should be, in particular, applicable to the bulk of QHE systems.  
For the case $d=\alpha=\beta=3$ the temperature scaling is $\kappa \propto T^{4/3}$. 

The spectral diffusion mechanism leads to the relaxation of thermal spins with the rate (\ref{tau_sd}). In the case of $\alpha=\beta=3$  the rate is given by Eqs. (\ref{tau_sd3}) and (\ref{tau_sd2}) in three and two spatial dimensions, respectively. 

We have further studied the thermal transport in the system in the regime $\beta/2<d<\alpha\beta/(\alpha+\beta)$ (region III in Fig.~\ref{Fig:PhaseDiagram}), where pseudo-spin resonances are not sufficient to delocalize the excitations, and the
delocalization of energy occurs via interaction of pseudo$^2$-spins.  In this case, the temperatures scaling of thermal conductivity is predicted to be given by Eq. (\ref{kappa2}) if the spectral diffusion is disregarded and by Eq. (\ref{kappa_2_sd}) with the spectral diffusion properly taken into account. 

In the region II$_{\rm c}$, the pseudo-spin transport mechanism coexists with the pseudo$^2$-spin one, so that the corresponding contributions to the thermal conductivity compete. Furthermore, in a part of this region close to the border with region III, the relaxation rate of thermal pseudo-spins is controlled by delocalized pseudo$^2$-spins. While we expect that this mixed mechanism dominates the thermal transport in a part of the region II$_{\rm c}$ near the border with III, we have not evaluated the corresponding contribution to thermal conductivity, leaving this as a prospect for future work. 

We have also determined the scaling of the mobility edge $E_c$ for many-body excitations with the system size $L$. The result is given by Eq.~(\ref{Ec}) in the region II$_{\rm b}$ and by Eqs.~(\ref{Ec2}), (\ref{gamma2}) in the regions II$_{\rm c}$ and III. The corresponding exponents $\gamma$ und $\gamma_2$, given by Eqs.~(\ref{gamma}) and (\ref{gamma2}), respectively, are intermediate between the cases  of Fock-space localization in a quantum dot, $\gamma=-d$ , and the many-body localization transition induced by a short-range interaction, $\gamma=d$. 

We conclude the paper by reviewing some further implications of our work and related research prospects; the work in these directions is currently underway.

\begin{enumerate}

\item[(i)]
 Our theory should be relevant not only to systems of localized electrons with Coulomb interaction but also to other realizations of the spin Hamiltonian (\ref{Hamiltonian}). As has been already discussed in Sec.~\ref{section-Resonant-spin-networks}, these include, in particular, interacting two-level systems in amorphous materials, as well as ensembles of dipole molecules in optical lattices and of solid-state spin defects. Let us emphasize that we have considered here only an ensemble of interacting spins, discarding all other degrees of freedom. In other words, we assumed that these other degrees of freedom are irrelevant for transport properties. Clearly, this is not necessarily the case. In particular, if spins represent atomic two-level systems, one may need to explore an interplay between phonons and the interacting spin system. Implications of our work for this situation remain to be explored.

 \item[(ii)]  
 It would also be interesting to extend our analysis onto the remaining part of the phase diagram, $\beta>\alpha>d$. The boundary between the delocalized and many-body localized phases in this part of the phase diagram was recently established in Ref.~\onlinecite{Burin15a}. The random XY model is a prominent representative of this class of models. A related problem is the random Ising model in transverse field. Effective spin models with long-range interactions of this class, which are of particular relevance in the context of cold atomic gases and Josephson junction arrays, have been recently considered in the literature \cite{Monroe13,Monroe15,Hauke_Heyl,Kettemann,Monthus}.

\item[(iii)] There is a large-body of experimental data that indicate that the electrons in a 2D system (with Coulomb interaction) deep in the Anderson-insulator regime can thermalize in the absence of phonon bath.
In particular, the prefactor in the hopping conductivity is of the order of $e^2/h$, see Refs.~\onlinecite{Mason95,Lam97,Simmons98,Khondacker99,Yakimov00}, which suggests a phononless mechanism of transport. (In the case of phonon-assisted  hopping, the prefactor would be much smaller and non-universal.)  Also, far-from-equilibrium measurements \cite{McCammon05} indicate that the electronic subsystem in the hopping-conductivity regime may form a thermal state characterized by a temperature strongly differing from the phonon temperature.  A development of consistent theory of these effects represents a major long-standing challenge.  One may expect that the delocalized pseudo-spin subsystem would work as a bath for electrons, thus providing a mechanism for the phononless transport. We thus hope that our work will pave the way for a development of a systematic theory of electron thermalization and phononless hopping in Anderson insulators. 

\item[(iv)] 
In Sec.~\ref{section-Localization-threshold} we have studied the localization threshold $E_c$ (or $T_c$) in a system of finite size $L$. It is worth emphasizing that the obtained result represent the threshold for the ``most mobile'' excitations -- spins and pseudospins  for the cases of delocalization via pseudo-spin network and pseudo$^2$-spin network, respectively. Other degrees of freedom (such as spins in the regions III and IIc of the phase diagram as well as electrons in all cases) are expected to become delocalized only at higher energies. Investigation of this hierarchy of delocalization thresholds represents an interesting research prospect.

 \end{enumerate}

 \section{Acknowledgment}

We acknowledge useful discussion with I. Burmistrov, M. Heiblum, K. Michaeli,  M. M\"uller, 
F. Pierre, D.G. Polyakov, and B.I. Shklovskii.
 This work was supported by ISF (grant 584/14),
by GIF (grant 1167-165.14/2011), and by Russian Science Foundation under the grant No. 14-42-00044 (I.V.G., I.V.P., and A.D.M.)

\appendix

\section{Matrix element}
\label{matrix-element}

In this Appendix we estimate the matrix element  of the interaction-induced process which corresponds to flipping three pseudo-spins. This process arises in the second order of perturbation theory (see also Ref.~\onlinecite{Burin07} for a similar calculation). 
As discussed in the main text, the pseudo-spin Hamiltonian after a unitary transformation (that orients all spins in $z$ direction) takes the form
(\ref{rotated}).
The transition rate between initial and final states is determined by golden rule
\begin{equation}
R_{fi}=|A_{fi}^{(2)}|^2\delta(E_i-E_f)\,,
\end{equation}
where the second order transition amplitude is given by
\begin{equation}
A_{fi}^{(2)}=\sum_m\frac{\langle f|U| m\rangle\langle m|U|i\rangle}{E_m-E_i}\,.
\label{Afi}
\end{equation}
We assume that in the initial state the spin 1 is oriented up, while the spins 2 and 3 are down. In the final state all spins are flipped. 
 The  energies of pseudo-spins are $\epsilon_1=\epsilon_2+\epsilon_3$,  $\epsilon_2$, and $\epsilon_3$.  
Calculating the matrix elements entering (\ref{Afi}) induced by the pseudo-spin interaction defined by Eq.(\ref{rotated}) and summing over the allowed intermediate states, one finds
\begin{eqnarray}&&
A_{fi}^{(2)}=2n_x^1n_x^2n_x^3   \nonumber\\ &&
\times\bigg[n_z^{(1)} u_{13}u_{12}\frac{\epsilon_1}{\epsilon_2\epsilon_3}-n_z^{(2)}u_{12}u_{23}\frac{\epsilon_2}{\epsilon_1\epsilon_3}-n_z^{(3)}u_{13}u_{23}\frac{\epsilon_3}{\epsilon_1\epsilon_2}\bigg]. \nonumber
\\
\label{Afi2}
\end{eqnarray} 
Since orientations of vectors $\vec{n}$ and values of matrix elements $u_{ij}$ are random, there is no reason to expect that cancellations between the terms may change the estimate for $A_{fi}^{(2)}$. Estimating the terms entering Eq.~(\ref{Afi2}) for the situation considered in Sec.~\ref{section-Thermal-transport-High-energy-excitations} (two high-energy pseudo-spins with a small energy difference $\omega$) , we come to the result presented in Eq.~(\ref{matrix-element-high-energy}).  \\
 
\section{Decay of a high-energy pseudo-spin into  two  pseudo-spins of approximately equal energy}
\label{decay}

In this Appendix, we estimate the decay rate of a high-energy pseudo-spin due to processes of the type shown in Fig.~\ref{fig3}. As the calculation shows, these processes yield a contribution which is much smaller than that given by processes shown in Fig.~\ref{fig4} and analyzed in the main text. Thus, this channel of decay plays no role for our results for thermal conductivity. Nevertheless, we present this calculation in an Appendix for the sake of completeness. 

To evaluate the rate of this decay process, it is important to keep in mind that $E/2$ states are also nearly localized, with a broadening of the levels that is much smaller that the mean level spacing.
($\Delta(E)\tau(E) \ll 1$).  The transition rate is therefore 
\begin{eqnarray}
\label{rate}
\frac{1}{\tau(E)}=\frac{A^2(E)}{\Delta^2(E/2)\tau(E/2)}\,.
\end{eqnarray}
The corresponding matrix element reads
\begin{equation}
A(E) =t^2V^2\left(\frac{T\rho^2}{W^2}\right)^\frac{2\alpha}{d}E^{\frac{2\alpha}{d}-3}\,.
\end{equation}
To determine $\tau(E)$, one needs to iterate Eq.(\ref{rate})
 $\ln_2(E/E_*)$  times, until the optimal network  with ``strongly delocalized'' states forming a conventional continuum ($\tau(E_*) \sim E_*^{-1}$) is reached.  One thus finds
\begin{equation}
\tau(E) \sim \frac{1}{E_*}\exp\left(\ln^2(E/E_*)\right).
\end{equation}  
Thus, the rate of such processes decreases with energy increasing $E$ faster than any power law and can thus be discarded \cite{Doublon}.

\widetext
\vspace{2cm}
\begin{center}
\textbf{\large Supplemental Material to ``Energy transport in the Anderson insulator''}
\end{center}
\setcounter{equation}{0}
\setcounter{figure}{0}
\setcounter{table}{0}
\setcounter{page}{1}
\setcounter{section}{0}
\makeatletter
\newcommand{\mysection}[1]{
 \stepcounter{section}
  \section*{ {\bf S-\Alph{section}. } #1} }

\renewcommand{\theequation}{S-\Alph{section}.\arabic{equation}}
\renewcommand{\thefigure}{S\arabic{figure}}
\renewcommand{\bibnumfmt}[1]{[S#1]}
\renewcommand{\citenumfont}[1]{S#1}

In this Supporting Material we provide some technical details of calculations for (i) spin relaxation rate and (ii) comparison of different contributions to the spin relaxation rate and the thermal conductivity.

\mysection{Spin relaxation due to interaction with the optimal network (neglecting spectral diffusion)}

\label{S-s1}

In this Section we present details of calculation of relaxation of typical thermal spins assisted by their interaction with resonant spin pairs (pseudospins) forming the optimal network. Within this calculation, the spectral diffusion is discarded.

The process is analogous to the flip-flop process of high-energy pseudo-spins assisted by optimal network, as analyzed  in Sec. VI of the article.
The relaxation channel we discuss here  is mentioned in the beginning of Sec. IX of the article. As stated there, the spectral diffusion provides in fact a more efficient channel of spin relaxation, see Sec. \ref{S-s2} of this Supplemental Material for a detailed comparison.

Let us assume that we are in the region II$_b$ of the phase diagram, that is $d < \beta<\alpha<2d$. 
We take two spins $s_1$ and $s_2$ at distance $r_{12}$ and allow them to interact with the optimal network of pseudospins $\tau$. The interaction Hamiltonian reads (before the rotation of $\tau$ spins to the eigenbasis)
\begin{equation}
H=\epsilon_i \tau^z_i+\Delta_i\tau^x_i+\frac{u_{ij}}{r_{ij}^\beta}\tau_z^i\tau_z^j+\left(\frac{u_{1i}}{r_{1i}^\beta}s_1^z+ \frac{u_{2i}}{r_{2i}^\beta}s_2^z\right)\tau_i^z+\frac{t_{12}}{r_{12}^\alpha}\left(s_1^+s_2^-+\ldots\right)+E_1s_1^z+E_2s^z_2.
\end{equation}
The z-z interaction of spins $s_1$ and $s_2$ is not important and we drop it. 

We rotate $\tau$ pseudo spins to the eigenbasis and compute the matrix element for the transition $$(s_1=\uparrow, s_2=\downarrow , \tau=\uparrow)\longrightarrow (s_1=\downarrow, s_2=\uparrow , \tau=\downarrow).$$
It is given by (on the mass shell)
\begin{equation}
A=\frac{t_{12} n_x}{r_{12}^\alpha}\left(\frac{u_{1\tau}}{r_{1\tau}^\beta}-\frac{u_{2\tau}}{r_{2\tau}^\beta}\right)\frac{1}{E_1-E_2}.
\end{equation}
Here $r_{1\tau}$  is the distance between $s_1$ and pseudospin  $\tau$. 
We demand $E_1-E_2\sim E_*$, which is required by the energy conservation. (Here $E_*$ is the characteristic energy on the optimal network.) Thus, we get the following equation for $r_{12}$: 
\begin{equation}
\rho r_{12}^d\frac{E_*}{W}\sim 1\,, \qquad r_{12}=\left(\frac{W}{\rho E_*}\right)^{1/d}.
\end{equation} 
The typical distance to a spin of the optimal network is 
$$r_{1\tau}\sim R_2^*(R_1(T))=\left[\frac{V R_1^\alpha(T)}{t}\right]^{1/\beta}.$$
Thus, we get the following estimate for the matrix element:
\begin{equation}
A\sim \frac{t V}{E_* R_2^\beta(T) r_{12}^{\alpha}}=\frac{t}{r^\alpha_{12}}.
\end{equation}
Now we can compute the spin relaxation rate given by \cite{S-Note1}
\begin{equation}
\frac1\tau\sim\frac{A^2}{E_*}=\frac{t^2}{E_* r^{2\alpha}_{12}}=\frac{t^2\rho^{2\alpha/d}}{W^{2\alpha /d}}E_*^{\frac{1}{d}(2\alpha-d)}=t\left(\frac{t\rho}{W}\right)^{2\alpha/d}
\frac{1}{\left[R_1(T)\right]^{\frac{\alpha}{d}(2\alpha-d)}}.
\end{equation}
Here $R_1(T)$ is the pseudospin size in the optimal network,
\begin{equation}
\label{S-R1}
R_1(T) \sim \bigg[\frac{Tt\rho^2}{W^2}\left(\frac{V}{t}\right)^{\frac{d}{\beta}}\bigg]^{-\frac{\beta}{(\alpha+\beta)d-\alpha\beta}}.
\end{equation}
The above calculation is essentially identical to the one in Sec. VI of the paper (relaxation of thermal pseudospins), with the only difference that the density of pseudospins is replaced by the density of spins. Since the spin density is larger, the resulting rate is also higher.

For $\alpha=\beta$ we have 
\begin{equation}
\label{S-aa}
R_1(T) \sim \bigg[\frac{Tt\rho^2}{W^2}\bigg]^{-\frac{1}{2d-\alpha}}, \qquad \frac{1}{\tau}\sim T^{\frac{\alpha}{d}\frac{2\alpha-d}{2d-\alpha}}\,, \qquad
 r_{12}\sim R_1^{\alpha/d}(T)\sim T^{-\frac{\alpha}{d}\frac{1}{2d-\alpha}}
\end{equation}

To estimate the contribution of the above spin-relaxation processes to the thermal conductivity, we take into account that the typical energy transfer is $T$ and the density of states for the typical thermal spins  is $\nu\sim \rho(T/W)/T\sim\rho/W$:
\begin{equation}
\kappa\sim\frac{T^3}{T^2}\nu \frac{r_{12}^2}{\tau}\sim \frac{t\rho T}{W} \left(\frac{t\rho}{W}\right)^{2\alpha/d}
\frac{1}{\left[R_1(T)\right]^{\frac{\alpha}{d}(2\alpha-d)}}\left(\frac{W}{t\rho}\right)^{2/d}R_1^{2\alpha/d}(T),
\end{equation}
which yields
\begin{equation}
\kappa \sim T  \left(\frac{t\rho}{W}\right)^{\frac{2\alpha-2+d}{d}}
\left[R_1(T)\right]^{-\frac{\alpha}{d}(2\alpha-d-2)}.
\end{equation}
Here $R_1(T)$ is given by Eq.~(\ref{S-R1}). This is valid in the whole range of $\alpha$ and $\beta$ as specified in the beginning of  this Section. 

In particular, in the case $\alpha=\beta$, we get for the thermal conductivity
\begin{equation}
\kappa\sim T^{1+\frac{\alpha}{d}\frac{(2\alpha-d-2)}{2d-\alpha}}
\end{equation}
For $\alpha=\beta=3$ and $d=3$ we get $\kappa \sim T^{4/3}$, while for $d=2$ we get $\kappa\sim T^4$.

\mysection{Comparison of contributions to spin relaxation rate and thermal conductivity}
\label{S-s2}
Here we collect formulas for scaling of relaxation rate and of the thermal conductivity within the (i) pseudospin calculation, Sec. IV of the article, (ii) ``conservative'' spin calculation (neglecting spectral diffusion, Sec.~{\color{red} A} of this Supplemental Material), and (iii) spin diffusion, Sec. IX of the article. We keep the scaling with the disorder strength and with temperature. For simplicity, we restrict ourselves here to the case $\alpha=\beta$ and $V=t$; it is easy to write down generalizations for arbitrary $\alpha$ and $\beta$.

\subsection{Relaxation rate}

Pseudospins (ps) at thermal energy:
\begin{equation}
1/\tau_{ps} \propto T^{\frac{5\alpha - 4d}{2d-\alpha}} W^{-\frac{6\alpha}{2d-\alpha}}
\end{equation}

Spins (s):
\begin{equation}
1/\tau_{s} \propto T^{\frac{\alpha(2\alpha - d)}{d(2d-\alpha)}} W^{-\frac{2\alpha(d+\alpha)}{d(2d-\alpha)}}
\end{equation}

Spins -- spectral diffusion (sd):
\begin{equation}
1/\tau_{sd} \propto T^{\frac{2\alpha - d}{2d-\alpha}} W^{-\frac{3\alpha}{2d-\alpha}}
\end{equation}
 
In particular, for $\alpha =\beta =3$ and $d=3$:
 \begin{equation}
1/\tau_{ps} \propto TW^{-6}, \qquad 1/\tau_{s} \propto TW^{-4},   \qquad 1/\tau_{sd} \propto TW^{-3}.
\end{equation}
For $\alpha =\beta =3$ and $d=2$:
 \begin{equation}
1/\tau_{ps} \propto T^7W^{-18}, \qquad 1/\tau_{s} \propto T^6W^{-15},   \qquad 1/\tau_{sd} \propto T^4W^{-9}.
\end{equation}

\subsection{Thermal conductivity}

Pseudospins (ps):
\begin{equation}
\kappa_{ps} \propto T^{\frac{(d+\alpha-2)(3\alpha-2d)}{\alpha(2d-\alpha)}}W^{-4\frac{\alpha+d-2}{2d-\alpha}}
\end{equation}

Spins (s):
\begin{equation}
\kappa_{s} \propto T^{1+\frac{\alpha}{d}\frac{(2\alpha-d-2)}{2d-\alpha}}W^{-\frac{2\alpha^2+(d-2)(2d+\alpha)}{d(2d-\alpha)}}
\end{equation}

Spins -- spectral diffusion (sd):
\begin{equation}
\kappa_{sd} \propto T^{\frac{(\alpha+d-2)}{2d-\alpha}}W^{-\frac{2(\alpha+d-2)}{2d-\alpha}}
\end{equation}
 
In particular, for $\alpha =\beta =3$ and $d=3$:
 \begin{equation}
\kappa_{ps} \propto T^{4/3}W^{-\rm 16/3}, \qquad \kappa_{s} \propto T^{4/3}W^{-3},   \qquad \kappa_{sd} \propto T^{4/3}W^{-8/3}.
\end{equation}
For $\alpha =\beta =3$ and $d=2$:
 \begin{equation}
\kappa_{ps} \propto T^5W^{-12}, \qquad \kappa_{s} \propto T^4W^{-9},   \qquad \kappa_{sd} \propto T^3W^{-6}.
\end{equation}

Thus, taking into account the spectral diffusion yields the most efficient mechanism of the spin relaxation and of the thermal transport.


\begin{thebibliography}{99}
\bibitem{footnote-dipole-dos}
The characteristic size of particle-hole dipoles $\xi_d$ with energy splittings $\epsilon_i \sim T$ is expected to be larger than the localization length $\xi$ by a logarithmic factor $\sim \log(1/T)$. While $\xi_d$ does not enter explicitly our results, it will enter, through the Coulomb energy $e^2/\xi_d$, the expression for the density of dipole excitations obtained in the framework of the Coulomb-gap theory \cite{SE}. This would lead to additional logarithmic-in-$T$ factors in the electronic model. We discard such factors in our analysis of temperature dependence of thermal conductivity, assuming a constant and temperature-independent density of states of dipole excitations.
\bibitem{note-optimal-pseudospin-size}  Equation (\ref{R1}) is consistent with related results on localization threshold in the infinite-temperature limit of the model presented in Table II of Ref.~\onlinecite{Burin15a}. See also footnote 29 of Ref.~\onlinecite{Burin15a} on matching of infinite-temperature and low-temperature limits.
\bibitem{Burin15} 
A related discussion of the scaling of critical disorder can be found in Ref.~\onlinecite{Burin15b}.
\bibitem{FootNoteLogBette} When estimating the localization threshold, we discard possible logarithmic factors that may emerge in
many-body problems in analogy with Anderson localization on a Bethe lattice with large connectivity, see
Refs. \onlinecite{gornyi05, basko06,  AGKL, mirlin97}.
\end{thebibliography}
\end{document}